\documentclass[twocolumn, 10pt]{IEEEtran}

\usepackage{hyperref, bm,graphicx,amsmath,amssymb}
\usepackage{algorithm,algpseudocode}

\usepackage{color, soul}

\begin{document}
\title{ Detection-Directed Sparse Estimation using Bayesian Hypothesis Test and Belief Propagation}

%\author{\IEEEauthorblockN{Jaewook Kang}}

\author{ Jaewook~Kang,~\IEEEmembership{Student Member,~IEEE,}
        Heung-No~Lee,~\IEEEmembership{Member,~IEEE,}
        and~Kiseon~Kim,~\IEEEmembership{Senior Member,~IEEE}\\
\bigskip
\IEEEauthorblockA{School of Information and Communication,\\
 Department of Nanobio Materials and Electronics,\\
 Gwangju Institute of Science and Technology (GIST), Gwangju, Republic of
 Korea}
 \vspace{-20pt}

% Ack
}
\maketitle \vspace{-20pt}

\setlength{\baselineskip}{0.93\baselineskip}
\begin{abstract}
In this paper, we propose a sparse recovery algorithm called
detection-directed (DD) sparse estimation using Bayesian hypothesis
test (BHT) and belief propagation (BP). In this framework, we
consider the use of sparse-binary sensing matrices which has the
tree-like property and the sampled-message approach for the
implementation of BP.

The key idea behind the proposed algorithm is that the recovery
takes DD-estimation structure consisting of two parts: support
detection and signal value estimation. BP and  BHT  perform the
support detection, and an MMSE estimator  finds the signal values
using the detected support set. The proposed algorithm provides
noise-robustness against measurement noise beyond the conventional
MAP approach, as well as a solution to remove quantization effect by
the sampled-message based BP independently of memory size for the
message sampling.

We explain how the proposed algorithm can have the aforementioned
characteristics via exemplary discussion. In addition, our
experiments validate such superiority of the proposed algorithm,
compared to recent algorithms under noisy setup. Interestingly the
experimental results show that  performance of the proposed
algorithm approaches that of the oracle estimator as SNR becomes
higher.
\end{abstract}


\section{Introduction}

Sparse signal recovery in the presence of noise has been intensively
investigated in many recent literature because any real-world
devices are subject to at least a small amount of noise. We refer to
such problems as \emph{noisy sparse signal recovery} (NSR) problems.
Let $\mathbf{x}_0 \in \mathbb{R}^N$ denote a sparse signal vector
whose elements are sparsely non-zeros. Then, the NSR decoder
observes a measurement vector $\mathbf{z}=\mathbf{\Phi}\mathbf{x}_0
+ \mathbf{n} \in \mathbb{R}^M$, where $\mathbf{\Phi} \in
\mathbb{R}^{M \times N}$ is a fat sensing matrix $(M \leq N)$; and
we limit our discussion to zero-mean independently identically
distributed (i.i.d.) Gaussian noise denoted by $\mathbf{n} \in
\mathbb{R}^M$.

The NSR problem can then be defined as an $l_0$-norm minimization
problem as similarly done in
\cite{Donoho2},\cite{tropp},\cite{Donoho1},\cite{Bruckstein}:
\begin{eqnarray}\label{eq:eq_2}
\widehat{\bf{x}}_{0,l_0} = \arg \mathop {\min }\limits_{} \left\|
{\bf{x}} \right\|_0 \,\,\,\,\,\,s.t.\,\,  \left\| {{\bf{\Phi
}}\mathbf{x} - {\bf{z}}} \right\|_2^2 \le \epsilon ,
\end{eqnarray}
where $\epsilon$ is an error tolerance parameter. In general, the
minimization task in \eqref{eq:eq_2} is NP hard; therefore,
$l_1$-norm approaches have been developed and discussed as
alternatives
\cite{Donoho2}-\cite{Lasso}. %, and it was shown that the
%$l_1$-solvers are equivalent to the $l_0$-solver under a relaxed
%condition \cite{Donoho1}.
Among the $l_1$-solvers, the Dantzig selector (L1-DS), proposed by
Candes and Tao \cite{candes2}, and the $l_1$-penalized least-square
algorithm usually called LASSO \cite{Lasso} has been devised for the
Gaussian noise setup.

Bayesian approaches to NSR have also received attention, where the
minimization task in \eqref{eq:eq_2} is  described as the maximum a
posteriori (MAP) problem, and the sparse solution then is sought by
imposing sparsifying prior density on $\mathbf{x}$ as follows
\cite{Bruckstein}-\cite{BCS-LP}:
\begin{eqnarray} \label{eq:eq_3}
\begin{array}{l}
\mathbf{\widehat{x}}_{0,\text{MAP}}=\arg\mathop {\max
}\limits_{\mathbf{x}} \,\,f_{\mathbf{X}}( \mathbf{x} |
\mathbf{Z}=\mathbf{z})\\
\,\,\,\,\,\,\,\,\,\,\,\,\,\,\,\,\,\,\,= \arg\mathop {\max
}\limits_{\mathbf{x}} \,\,f_{\mathbf{Z}}( \mathbf{z} |
\mathbf{X}=\mathbf{x})f_{\mathbf{X}}( \mathbf{x})
\end{array}
\end{eqnarray}
where  $f(\cdot)$ is a probability density function (PDF).

The task in \eqref{eq:eq_3} is likewise computationally demanding;
hence, many relaxed-Bayesian solvers have been devised according to
various prior types and applied techniques, such as sparse
Bayesian-learning (SBL) \cite{SBL}-\cite{MCMC},\cite{BP-SBL},
Approximate minimum-mean-squared-error (AMMSE) estimation
\cite{MMSE-CS2}-\cite{JWKANG_SPARS11}, and belief propagation (BP)
with sparse sensing matrices (BP-SM) \cite{CS-BP1}-\cite{SuPrEM}. A
summary of the relaxed-Bayesian solvers is in Table \ref{table1}.

\begin{table*}[!t]\label{table1}
\renewcommand{\arraystretch}{1.0}

\caption{Relaxd-Bayesian solvers for sparse estimation }
\label{table1}
 \centering
\begin{tabular}{||c||c|c|c||}
\hline\hline
Algorithm type &  Sensing matrix type   & Prior type                            & Utilized techniques \\
\hline \hline
BP-SM  &      Sparse-ternary \cite{CS-BP1},\cite{CS-BP2}      & Gaussian-mixture \cite{CS-BP1},\cite{CS-BP2}        & Belief-Propagation (BP) \cite{CS-BP1},\cite{CS-BP2}\\
       &      Sparse-binary \cite{BP-SBL},\cite{SuPrEM}      & Hierarchical-Gaussian \cite{BP-SBL},\cite{SuPrEM}      & Expectation-maximization (EM), BP \cite{BP-SBL},\cite{SuPrEM} \\
\hline
        &  Gaussian-random \cite{MMSE-CS2}   &         Spike-and-slab \cite{MMSE-CS2}                          & Fast-Bayesian-Matching-Pursuit \cite{MMSE-CS2} \\
AMMSE   &  Training-based  \cite{MMSE-CS3}  &          Gaussian-mixture \cite{MMSE-CS3},\cite{MMSE-CS4}       & Random-OMP \cite{MMSE-CS3} \\
        &  Random-unitary \cite{MMSE-CS4}   &                                                                    & Closed-form-MMSE \cite{MMSE-CS4}\\
\hline
        &   Gaussian-kernel  \cite{SBL}                 & Hierarchical-Gaussian  \cite{SBL}-\cite{BCS}   & Relevance-vector-machine, EM \cite{SBL}-\cite{BCS-LP}\\
   SBL  &   Gaussian-random  \cite{SBL-BS}-\cite{MCMC}     & Spike-and-slab \cite{MCMC}                            & Markov-chain-Monte-Carlo \cite{MCMC} \\
        &   Uniform-random   \cite{BCS-LP}                   & Hierarchical-Laplace \cite{BCS-LP}                       & \\
\hline\hline
\end{tabular}

\end{table*}

We are mainly interested in  the  BP-SM framework, in this paper,
which has been investigated as a low-computational approach to solve
linear estimation problems such as
$\mathbf{z}=\mathbf{\Phi}\mathbf{x}_0 + \mathbf{n}$.  In this
framework, the matrix $\mathbf{\Phi}$ is assumed to be a sparse
matrix which has the tree-like property, and the statistical
connection of $\mathbf{x}_0$ and $\mathbf{z}$ is described with the
bipartite graph model of $\mathbf{\Phi}$. The tree-like property
ensures that the corresponding graph is asymptotically cycle-free
\cite{Guo06},\cite{Guo07},\cite{Richardson}. In addition, it is
known that the vector finding problem can be decomposed to a
sequence of scalar finding problems in the BP-SM framework, where
marginal posterior for each scalar estimation is obtained by an
iterative message-passing process. This decomposition have been
explained by \emph{decoupling principle} \cite{Guo05},\cite{Guo07}.

For implementation of BP,  two approaches have been mainly discussed
according to the message representation: 1) the sampled-message
based BP \cite{non_para_BP},\cite{Noorshams} where the message is
directly sampled from the corresponding PDF with a certain step size
such that the message is treated as a vector, and 2) the
parametric-message based BP (also called relaxed BP)
\cite{Guo06},\cite{para_BP},\cite{rangan} where the message is
described as a function with a certain set of parameters such as
mean and variance. If the sampled-message approach is chosen,
quantization error will be induced according to the step size. If
the parametric-message approach is used, some model approximation
must be applied  for stable iteration at the expense of
approximation error.

As applications of the BP-SM approach, the \emph{low-density
parity-check} coding \cite{Gallager}-\cite{Richardson}  and the CDMA
multiuser detection problems
\cite{Guo05},\cite{Guo08},\cite{Tanaka05} are well known  in
addition to the NSR works \cite{CS-BP1}-\cite{SuPrEM}.

Based on the rigorous theoretical support for the BP-SM framework by
Guo \emph{et al.} \cite{Guo05}-\cite{Guo08} and Tanaka \emph{et al.}
\cite{Tanaka05}, and motivated by recent NSR works by Baron \emph{et
al.} \cite{CS-BP1},\cite{CS-BP2}, Tan \emph{et al.} \cite{BP-SBL},
and Akcakaya \emph{et al.} \cite{SuPrEM}, in this paper, we aim to
develop a BP-SM type algorithm as an alternative for solving
\eqref{eq:eq_3}.  We refer to the proposed algorithm as
\emph{Detection-directed sparse estimation via Bayesian hypothesis
test and belief propagation} (BHT-BP). Differently from the works
\cite{CS-BP1}-\cite{SuPrEM} solving the MAP problem in
\eqref{eq:eq_3}, the proposed algorithm takes a structure of
detection-directed (DD) estimation which consists of a signal
support detector and a signal value estimator. The support detector
is designed using a combination of the sampled-message based BP  and
a novel Bayesian hypothesis test (BHT), and the signal value
estimator behaves in minimum mean-square error (MMSE) sense. The
DD-structure considers the common procedure of first using the
measurements at hand to detect the signal support set. This detected
support is then used in the model of the sparse signal, and an MMSE
estimator finds the signal values as if the detected support set is
in fact the correct set.

This DD-methodology was originally investigated by Middleton
\emph{et al.} for estimation of noisy signals \cite{DE_info}, and
includes wide application areas, such as communication systems
\cite{Picchi87},\cite{Godard80} and speech processing
\cite{ephraim84}. For NSR, a line of  previous works in the AMMSE
group has independently studied similar structures by Schnitter
\emph{et al.}, Elad \emph{et al.}, and Lee in
\cite{MMSE-CS2}-\cite{JWKANG_SPARS11},\cite{HNLEE}.

Then, the proposed algorithm achieves the following properties:
\begin{enumerate}
\item Providing robust signal support detection against the measurement
noise,
\item Removing quantization effect caused by the use of the sampled-message
based BP.
\end{enumerate}
Here, the ``oracle estimator" implies the estimator which has the
perfect support knowledge.

The combination of BHT and BP enables robust detection of the signal
support against measurement noise, which was partially introduced in
our previous work \cite{SSP2012}. In the support detection, the
sampled-message based BP provides marginal posterior for each scalar
problem according to the decoupling principle, and the BHT-process
then detects the supportive state of each scalar element  by
measuring the inner products between the marginal posterior and
reference functions composed of the prior density. This BHT-detector
utilizes the signal posterior information more efficiently than the
conventional MAP-approaches \cite{CS-BP1}-\cite{SuPrEM}. When the
measurements are noisy, density spread occurs in  marginal
posteriors, leading difficulty in making correct decisions via the
MAP-approach. In contrast, the BHT-based support detector
compensates such a weakpoint by scanning the entire range of the
posterior. Such hypothesis-based support detectors have been
discussed in wavelet domain denoising problems
\cite{Pizurica06},\cite{Vidakovic98}; however, they were using
thresholding techniques to sort out significant wavelet
coefficients, different from our work which uses the inner product.

In addition, we emphasize that the proposed algorithm effectively
removes quantization effect coming from the use of the
sampled-message based BP. The quantization effect limits performance
of the MAP-approach in both the signal value estimation and the
support detection. In the proposed algorithm, the use of the
DD-structure makes the performance independent of the message
sampling in the BP. In addition, we eliminate the quantization error
in the support detection by applying the minimum value of the signal
on its support, denoted by $x_{min}$, to the reference functions.
The importance of $x_{min}$ in the NSR problems was theoretically
highlighted by Wainwright \emph{et al.} in
\cite{Wainwright1},\cite{Wainwright2} where they showed that the
perfect support recovery is very difficult even with arbitrarily
large SNR if $x_{min}$ is very small. Hence, we regulate $x_{min}$
in our signal model, reflecting the knowledge of $x_{min}$  in the
BHT-detection. To the best of our knowledge, we have not seen the
reflection of $x_{min}$ in practical algorithms in the sparse
recovery literature, and surprisingly, this reflection enables
performance of the proposed algorithm  to approach that of the
oracle estimator as SNR increases.

 The computational complexity of the proposed algorithm is $O(N\log N + KM)$
 (if the cardinality of the support is fixed to
 $K$), which includes an additional cost $O(KM)$ owing to the BHT-process to the cost of BP $O(N
\log N)$. Nevertheless, the  cost of the proposed algorithm is lower
in practice since the proposed algorithm can catch the signal
support with  smaller memory size than NSR algorithms only using the
sampled-message based BP.

The remainder of the paper is organized as follows. We first briefly
review a line of the BP-SM algorithms, and then make a remark for
the relation between the BP-SM solvers and approximate message
passing (AMP)-type algorithms in Section II. In Section III, we
define our problem formulation. Section IV provides precise
description for the proposed algorithm and Section V provides
exemplary discussion to explain and support strengths of  the
proposed algorithm. In Section VI, we provide experimental
validation to show the advantage of the proposed algorithm and
compare to the recent algorithms. Finally, we conclude the paper in
Section VII.

\section{Related Works}
In this section, we provide a brief introduction to the previous
works in the BP-SM algorithms \cite{CS-BP1}-\cite{SuPrEM}.  The
common feature in these algorithms is the use of BP in conjunction
with sparse sensing matrices, to approximate the signal posterior,
where a sparsifying prior is imposed according to the signal model.
In addition, we make a remark to distinguish the  BP-SM works  from
AMP-type algorithms.

\subsection{BP-SM Algorithms} Baron
\emph{et al.} for the first time proposed the use of BP to the
sparse recovery problem with sparse sensing matrices
\cite{CS-BP1},\cite{CS-BP2}. The algorithm is called CS-BP. Signal
model of CS-BP is a compressible signal which has a small number of
large elements and a large number of near-zero elements. The author
associated this signal model with two-state mixture Gaussian prior,
given as
\begin{eqnarray}\label{eq:eq_4}
{f_{\mathbf{X}}}(\mathbf{x}) = \prod\limits_{i = 1}^N{
q\mathcal{N}(x_i;0,\sigma _{{X_1}}^2) + (1 -
q)\mathcal{N}(x_i;0,\sigma _{{X_0}}^2) },
\end{eqnarray}
where $q \in [0,1)$ denotes the probability that an element has the
large value, and $\sigma_{X_1}  \gg \sigma_{X_0}$. Therefore, the
prior is fully parameterized with $\sigma_{X_0},\sigma_{X_1}$, and
$q$. CS-BP performs MAP or MMSE estimation using the signal
posterior obtained from BP, where the authors applied both the
sampled-message and the parametric-message approaches for the
BP-implementation. The recovery performance is not very good when
measurement noise is severe since the CS-BP was basically designed
under noiseless setup.

 Tan \emph{et al.} proposed an algorithm under the BP-SM setup called BP-SBL
\cite{BP-SBL}.
 This work is based on the SBL-framework
\cite{SBL}-\cite{BCS-LP} which uses two-layer hierarchical-Gaussian
prior models given as
\begin{eqnarray}\label{eq:eq_5}
{f_{\mathbf{X}}}({\bf{x}}|a,b) = \prod\limits_{i = 1}^N
{\int_0^\infty {\mathcal{N}({x}_i;0,\tau _i^{ - 1})}  f_{\Gamma}
({\tau_i}|a_i,b_i)d{\tau_i }},
\end{eqnarray}
where $f_{\Gamma} ({\tau }|a,b)$ is the hyper-prior following Gamma
distribution with its parameters $a_i,b_i$. At each iteration, the
parameters $a_i,b_i$ of the prior are estimated using expectation
maximization (EM). Therefore, the posterior for the signal
estimation is iteratively approximated from the prior. The authors
applied the BP-SM setup to reduce the computational cost of EM.
BP-SBL is an algorithm using parametric-message based BP where every
message is approximated to be a Gaussian PDF which can be fully
described by its mean and variance. In addition, BP-SBL is input
parameter-free, which means this algorithm is adaptive to any signal
models and noise level since EM estimates the parameters associated
with any given models during the iteration. However, such parameter
estimation will not be accurate when noise is severe; therefore,
denoising ability of BP-SBL is limited when measurements are highly
corrupted.

 Most recently, Akcakaya \emph{et al.} proposed SuPrEM under a  framework
 similar to BP-SBL which uses a combination of EM and
the parametric-message based BP  \cite{SuPrEM}. The main difference
from BP-SBL is the use of a specific type of hyper-prior called
Jeffreys' prior $f_{\mathcal{J}}(\tau_i) =1/ \tau_i \,\, \forall
\tau \in [T_{i},\infty]$. The use of Jeffreys' prior reduces the
number of input parameters while sparsifying the signal.
 Therefore, the prior is given as
\begin{eqnarray}\label{eq:eq_5}
{f_{\mathbf{X}}}({\bf{x}}) = \prod\limits_{i = 1}^N {\int_0^\infty
{\mathcal{N}({x}_i;0,\tau _i)}  f_{\mathcal{J}} ({\tau_i})d{\tau_i
}}.
\end{eqnarray}
The sensing matrix used in SuPrEM is restricted to a sparse-binary
matrix which has fixed column and row weights, called low-density
frames. They are reminiscent of the regular LDPC codes
\cite{Gallager}. In addition, the signal model is confined to
$K$-sparse signals consisting of $K$ nonzeros and $N-K$ zeros since
SuPrEM includes a spasifying step which choose the $K$ largest
elements at each end of iteration. The noise statistic is an
optional input to the algorithm. Naturally, if the noise information
is available, SuPrEM will provide an improved recovery performance.

\begin{figure}[!t]
\centering
\includegraphics[width=8cm]{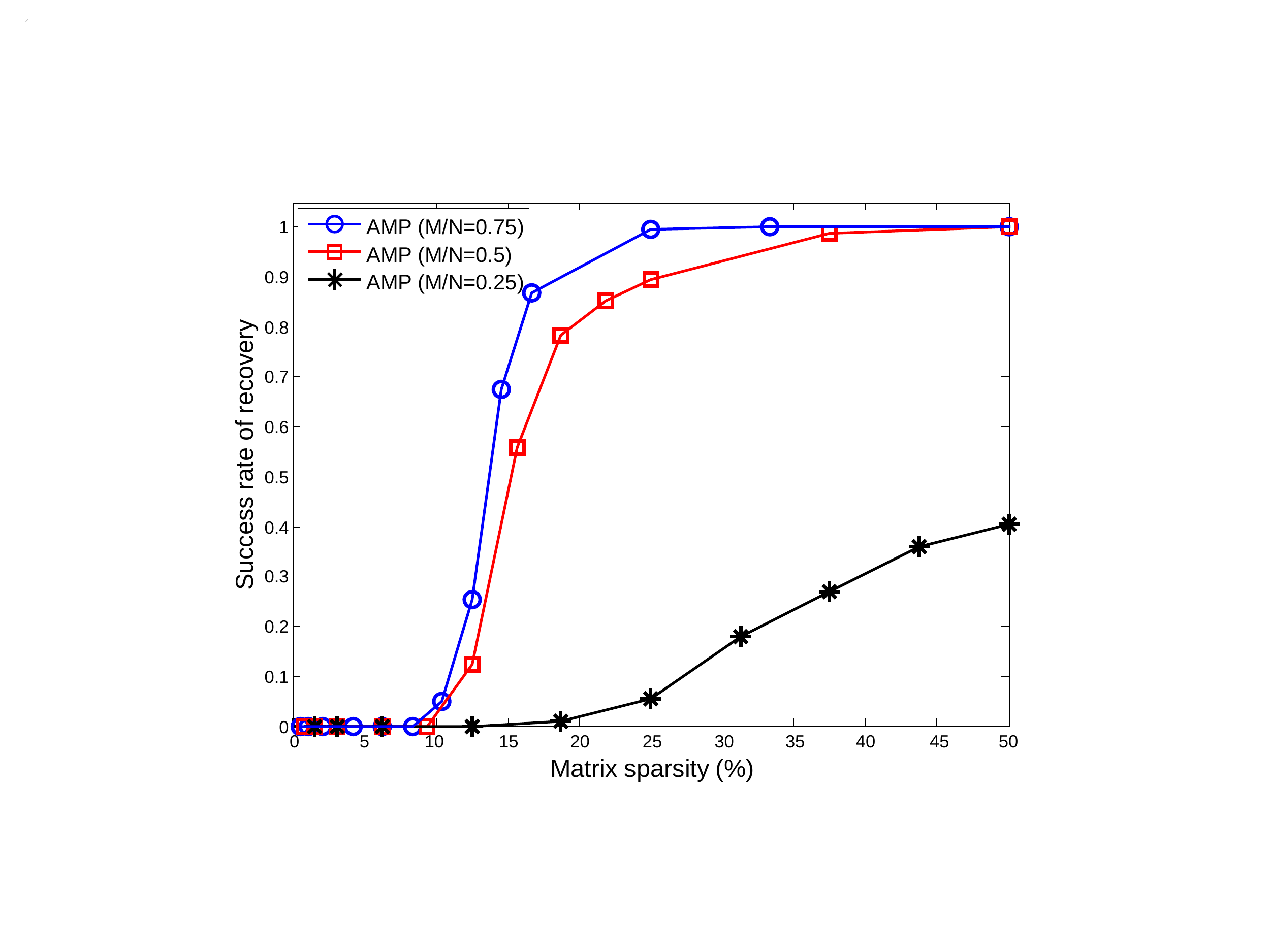}
\caption{Experimental success rate of sparse recovery via the
standard AMP under the use of sparse-Gaussian matrices. The success
rate is plotted as a function of the matrix sparsity  defined as
percentage of nonzero entries in the sensing matrix. We record a
success if the MSE, defined in \eqref{eq:eq3-9}, is below $10^{-3}$.
Reconstruction signal length is $N=1024$ and 500 iterations were
used.  } \label{fig:Fig2-2}
\end{figure}

\subsection{Relation to AMPs}
It is interesting to associate the algorithms in the BP-SM group to
the class of AMP algorithms, which was originally invented by Donoho
\emph{et al.} in \cite{AMP1},\cite{AMP2}, analyzed and refined by
Bayati \emph{et al.} \cite{AMP3} and Rangan \cite{GAMP}, because
both types of the algorithms utilize BP. Performance of the AMPs
coincides with that of the $l_1$-solvers under the large limit
assumption ($N \rightarrow \infty$) in the noiseless case
\cite{AMP1},\cite{AMP3}. Furthermore, in a recent result by Donoho
\emph{et al.} showed that performance of the AMPs is also equivalent
to LASSO \cite{Lasso} under the noisy setting with appropriate
parameter calibration \cite{NS_CS}.

The AMPs work well when sufficient density of the sensing matrix is
maintained as $N \rightarrow \infty$ since the AMPs were developed
on the basis of an approximation of BP-messages  by the central
limit theorem \cite{AMP2}. Therefore, if a sparse matrix is applied
to the AMPs, the  message approximation does not work, then the
decoder will fail in the recovery. We validate this claim by our own
experimental result shown in Fig.\ref{fig:Fig2-2}, where we
simulated the recovery success rate of the standard AMP \cite{AMP1}
without additive noise as a function of the matrix sparsity defined
as percentage of nonzero entries in the sensing matrix. The recovery
of the AMP are not successful when the sensing matrix is sparse (
typically less than 10$\%$ matrix sparsity) regardless of the number
of measurements $M$, as shown in an example of Fig.\ref{fig:Fig2-2}.
Namely, the AMPs recover sparse signals well with dense matrices at
the expense of low computation $O(MN \log N)$, but it does not enjoy
the benefits from the use of sparse matrices.

\section{Problem Formulation}
The goal of the proposed algorithm is to recover the object signal
$\widehat{\mathbf{x}}_0$ given the knowledge of $\mathbf{\Phi}$ and
noisy measurements $\mathbf{z}$ as following
\begin{eqnarray}\label{eq:eq3-4}
\mathbf{z}=\mathbf{\Phi}\mathbf{x}_0 + \mathbf{n},
\end{eqnarray}
where we consider the use of a fat sparse-binary sensing matrix
$\mathbf{\Phi} \in \{0,1\}^{M \times N}$ $(M<N)$ which has very low
matrix sparsity (typically less than 10$\%$ matrix sparsity) and
the tree-like property. We regulate the matrix sparsity using the
fixed column weight $L$ since this regulation enables the matrix
$\mathbf{\Phi}$ to span the measurement space with basis having
equal energy. For the noise distribution, we assume i.i.d. zero-mean
Gaussian density such that the noise vector $\mathbf{n} \in
\mathbb{R}^M$ is drawn from $
{\mathcal{N}(0,\sigma_N^2\mathbf{I})}$.

In the remainder of this section, we introduce some necessary
concepts, such as signal model and graph representation of
$\mathbf{\Phi}$, for our framework, and then discuss our approach to
solve this NSR problem.

\subsection {Graph Representation of $\mathbf{\Phi}$}
Bipartite graphs effectively represent linear systems with sparse
matrices such as the matrix $\mathbf{\Phi}$. Let
$\mathcal{V}:=\{1,...,N\}$ denote the set of indices corresponding
to the elements of the signal vector,
$\mathbf{x}_0=[x_{0,1},...,x_{0,N}]$, and $\mathcal{C}:=\{1,...,M\}$
denote the set of indices corresponding to the elements of the
measurement vector, $\mathbf{z}=[z_1,...,z_M]$. In addition, we
define the set of edges connecting $\mathcal{V}$ and $\mathcal{C}$
as $\mathcal{E} := \{ (j,i) \in \mathcal{V} \times
\mathcal{C}\,|\,\,\phi _{ji} = 1\}$ where $\phi _{ji}$ is the
$(j,i)$-th element of $\mathbf{\Phi }$. Then, a bipartite graph
$\mathcal{G}=(\mathcal{V,C,E})$ fully describes the neighboring
relation in the linear system. For convenience, we define the
neighbor set of $\mathcal{V}$ and $\mathcal{C}$ as $N_{\mathcal{V}}
(i) := \{ j \in \mathcal{C}\,|(j,i) \in {\mathcal{E}}\}$ and
$N_\mathcal{C} (j) := \{ i \in \mathcal{V}\,|(j,i) \in
{\mathcal{E}}\}$, respectively. Note that $\left|N_{\mathcal{V}}
(i)\right|=L$ under our setup on $\mathbf{\Phi}$.

\begin{figure}[!t]
\centering
\includegraphics[width=9cm]{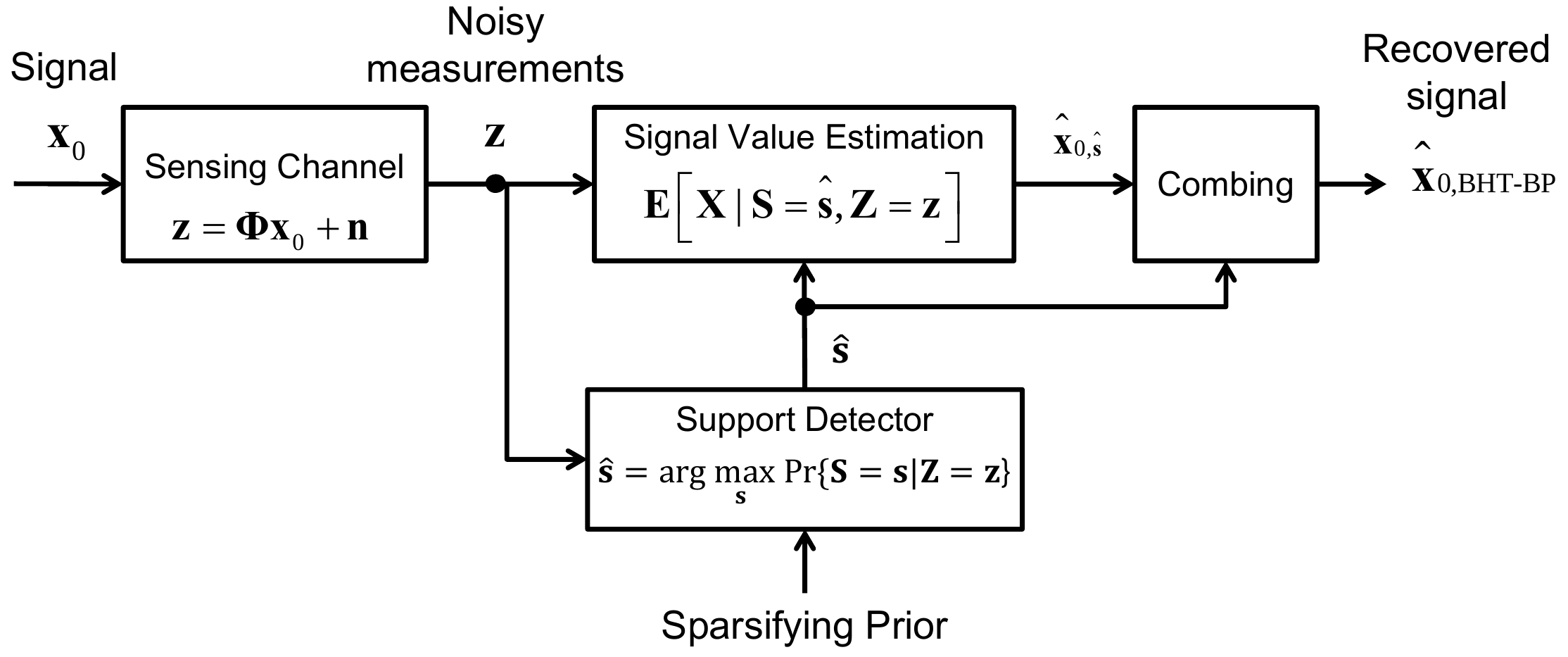}
\caption{DD-estimation in the proposed algorithm} \label{fig:Fig2-1}
\end{figure}

\subsection {Signal Model}
Let $\mathbf{x}_0 \in \mathbb{R}^N$ denote a sparse vector which is
a deterministic realization of a random vector $\mathbf{X}$. Here,
we assume that the elements of $\mathbf{X}$ are \emph{i.i.d.}, and
each $X_i$ belongs to the support set, denoted by
$supp(\mathbf{X})$, with a rate $q \in [0,1]$. To indicate the
supportive state of  $X_i$, we define a state vector
$\mathbf{S}(\mathbf{X})$ where each $S_i \in \mathbf{S}(\mathbf{X})$
is defined as
\begin{eqnarray}\label{eq:eq3-1}
{S_i}=S(X_{i}) = \left\{ \begin{array}{l}
1,\,\,\,\,\,{\rm{if}}\,\, i \in supp(\mathbf{X})\\
0,\,\,\,\,\,{\rm{else}}
\end{array} \right.\text{ for all }\,\, i \in \mathcal{V}.
\end{eqnarray}
Therfore, the signal sparsity is given as $K=||\mathbf{S}||_0$. For
the signal values on the support, we will deal with two cases, which
are
\begin{enumerate}
\item Gaussian signals:$\,X_{i \in supp(\mathbf{X})} \sim
\mathcal{N}(0,\sigma_{X_{1}}^2)$,
\item Signed signals  :$\,\,\,\,\,X_{i \in supp(\mathbf{X})} \sim
\frac{1}{2}\delta_{-\sigma_{X_{1}}}+\frac{1}{2}\delta_{+\sigma_{X_{1}}}$,
\end{enumerate}
where $\delta_{\tau}$ denote the delta function peaked at $\tau$.
Namely, in the first case, the values on the support are distributed
according to an i.i.d. zero-mean Gaussian density with variance
$\sigma_{X_{1}}^2$, and  in the second case the magnitude of the
values is fixed to $\sigma_{X_{1}}$ and the sign follows Bernoulli
distribution with probability $1/2$. In addition,  for the Gaussian
signal, we regulate the minimum value $x_{min}$ on the support as a
key parameter of the signal model, such that
\begin{eqnarray}\label{eq:eq3-2}
x_{min} \leq |x_{0,i}|\,\,\, \text{ for all } i \in
supp(\mathbf{x}_0),
\end{eqnarray}
Indeed, Wainwright \emph{et al.} emphasized the importance of
$x_{min}$ in the NSR problems \cite{Wainwright1},\cite{Wainwright2}.
They stated that if $x_{min}$ is very small, success of the noisy
sparse recovery is very difficult even with arbitrarily large SNR.
Namely, if the signal contains at least such a very small element
belonging to the support, the recovery can be failed regardless of
SNR level since the small element can be buried even by negligible
noise.

In the Bayesian framework, the prior density takes a role of
pursuing the sparsest solution from infinitely many solutions of the
underdetermined system.  Therefore, the proper choice of the
sparsifying prior according to the signal model is highly important.
According to our signal model, we consider \emph{spike-and-slab}
densities as our prior. The prior for an element $X_i$ is given as
\begin{eqnarray} \label{eq:eq4-1}
\begin{array}{l}
f_{X}(x)= qf_{X}( x |S = 1)+ (1 -
q)f_{X}( x |S = 0)\\
\,\,\,\,\,\,\,\,\,\,\,\,\,\,\,\,\,\,\,= q\mathcal{N}(x;0,\sigma
_{X_{1}}^2 )+ (1 - q)\delta_0.
\end{array}
\end{eqnarray}
This type of the priors has been widely used in modeling of strictly
sparse signals \cite{MCMC},\cite{MMSE-CS2},\cite{spikeandslab} since
the prior can
 simply characterize the signals with $\sigma_{X_{1}}$ and $q$, as well as easily
associate the signal model with the other models such as hidden
Markov chain and wavelet model. In addition, we note that the
spike-and-slab prior is a particular case of two-state Gaussian
mixture in \eqref{eq:eq_4} as $\sigma _{X_0} \rightarrow 0$.

\subsection {Solution Approach}
The approach of the DD-estimation is motivated by the naive MMSE
sparse estimation which was also reported in
\cite{MMSE-CS2}-\cite{MMSE-CS4}, as follows:
\begin{eqnarray}\label{eq:eq3-5}
{\widehat {\bf{x}}_{0,\text{MMSE}}} = \sum\limits_{{\bf{s}} \in {{\{
0,1\} }^N}} {{\bf{E}}[{{\bf{X}}}|{\bf{S}},{\bf{Z}}={\bf{z}}] \cdot
\Pr \{ {\bf{S}} |{\bf{Z}}={\bf{z}}\} \,},
\end{eqnarray}
We note that \eqref{eq:eq3-5} is a weighted sum of the signal value
estimation ${\bf{E}}[{{\bf{X}}}|{\bf{S}},{\bf{Z}}={\bf{z}}]$ over
$2^N$ possible supports given noisy measurements $\mathbf{z}$.
Therefore,  we separate the signal value estimation and  the $2^N$
support searching  in \eqref{eq:eq3-5} as the first step for its
relaxation. Then, the signal value estimation can be represented as
\begin{align}\label{eq:eq3-6-2}
&{\widehat {\bf{x}}_{0,\text{BHT-BP}}} = {\bf{E}}\left[
{{{\bf{X}}}|{\bf{S}} = {\widehat{\bf{s}}},{\bf{Z}}={\bf{z}}}
\right],
\end{align}
by assuming an ideal support detector for $\widehat {\bf{s}}$. The
calculation of \eqref{eq:eq3-6-2} is simple because it is  well
known as the convectional linear MMSE estimation,  expressed as
\begin{eqnarray}\label{eq:eq3-6-1}
{\widehat {\bf{x}}_{0,\widehat{\bf{s}} }} = {\left(
{\frac{1}{{\sigma_{X_{1}}^2}}{\bf{I}} + \frac{1}{{\sigma
_N^2}}{\bf{\Phi }}_{ \widehat{\bf{s}} }^T{\bf{\Phi }}_{
 \widehat{\bf{s}}}} \right)^{ - 1}}
\frac{1}{{\sigma _N^2}}{\bf{\Phi }}_{  \widehat{\bf{s}}}^T{\bf{z}},
\end{eqnarray}
where $\mathbf{\Phi}_{\widehat{\mathbf{s}}} \in \{0,1\}^{M \times
K}$ as a submatrix of $\mathbf{\Phi}$ that contains only the columns
corresponding to the detected support $\widehat{\mathbf{s}}$, and
$\mathbf{x}_{0,\widehat{\mathbf{s}}} \in \mathbb{R}^K$ as a vector
that includes only the nonzero elements from $\widehat
{\bf{x}}_{0}$. In addition, we know that this MMSE estimation is
optimal when the signal values on the detected support $\widehat
{\bf{s}}$ are Gaussian distributed \cite{KayI}.

For the part of the $2^N$ support search, we start from an
exhaustive detection, described as
\begin{align}\label{eq:eq3-6}
 &\widehat {\bf{s}} = \arg \mathop {\max
}\limits_{\mathbf{s} \in \{0,1\}^N} \Pr \{ {\bf{S}} =
{\bf{s}}|{\bf{Z}} = {\bf{z}}\}.
\end{align}
To solve \eqref{eq:eq3-6} efficiently, we decompose the state vector
detection  to $N$ scalar state detections  based on the decoupling
principle \cite{Guo05},\cite{Guo07}, and apply binary scalar
MAP-detection for each scalar state. Then, the binary scalar
detection   can be optimally  achieved by a hypothesis test
\cite{KayII}, given as
\begin{eqnarray}\label{eq:eq3-7}
{\frac{{\Pr \{  S_i = 1|{\bf{Z}}={\bf{z}}\} }}{{\Pr \{S_i =
0|{\bf{Z}}={\bf{z}}\} }}} \mathop {\mathop \gtrless \limits_{{H_0}}
}\limits^{{H_1}} 1 \text{ for all } i \in \mathcal{V},
\end{eqnarray}
where $H_0:S(X_{i})=0$ and $H_1:S(X_{i})=1$ are two possible
hypotheses. This support detection approach  simplifies the $2^N$
exhaustive search in \eqref{eq:eq3-6} to $N$ hypothesis tests. Here,
we note that the support detection in \eqref{eq:eq3-7} performs with
an independent decision rule from  the value estimation in
\eqref{eq:eq3-6-1}.  Therefore, this DD-approach to \eqref{eq:eq3-5}
is reasonable since the Bayesian risks of the support detection and
the signal values estimation can be independently minimized
\cite{DE_info}. The overall flow of the DD-estimation is depicted in
Fig.\ref{fig:Fig2-1}.

In order to justify our solution approach, we will provide
experimental validation compared to the recent NSR algorithms, as a
function of SNR defined as
\begin{eqnarray}\label{eq:eq3-10}
\text{SNR } :=10 \log_{10} \frac{{E{\left\| {{\bf{\Phi}}
\mathbf{x}_0} \right\|_2^2 } }}{{M\sigma _{N }^2 }} \text{ dB }.
\end{eqnarray}
Performance of the support detection is evaluated pairwise state
error rate (SER), defined as
\begin{eqnarray}\label{eq:eq3-8}
{\text{SER}}: = \Pr\{s(x_{0,i}) \ne \widehat{s}_i | i \in
\mathcal{V} \},
\end{eqnarray}
and the overall performance of the signal recovery is measured in
terms of normalized mean square error (MSE), defined as
\begin{eqnarray}\label{eq:eq3-9}
{\text{MSE}}: = \frac{ {\left\| {\widehat{\bf{x}}_0 - {\bf{x}}_0}
\right\|_2^2 } }{{\left\| {\bf{x}}_{0,\mathbf{s}} \right\|_2^2 }}.
\end{eqnarray}

\section{Proposed Algorithm}
The proposed algorithm straightforwardly provides the estimate
$\widehat{\mathbf{x}}_0$ via the MMSE estimation in
\eqref{eq:eq3-6-1} once given the detected support set
$\widehat{\mathbf{s}}$ from \eqref{eq:eq3-6}. In addition, we state
that the detected support set can be elementwisely obtained from
\eqref{eq:eq3-7}, on the basis of  the decoupling principle.
Therefore,  efficient design of  the scalar state detector in
\eqref{eq:eq3-7}  is very important in this work. In this section,
we explain the implementation of  the scalar state detector based on
the hypotesis test given in \eqref{eq:eq3-7}, using the combination
of the sampled-message based BP and BHT.

\begin{figure}[!t]
\centering
\includegraphics[width=9cm]{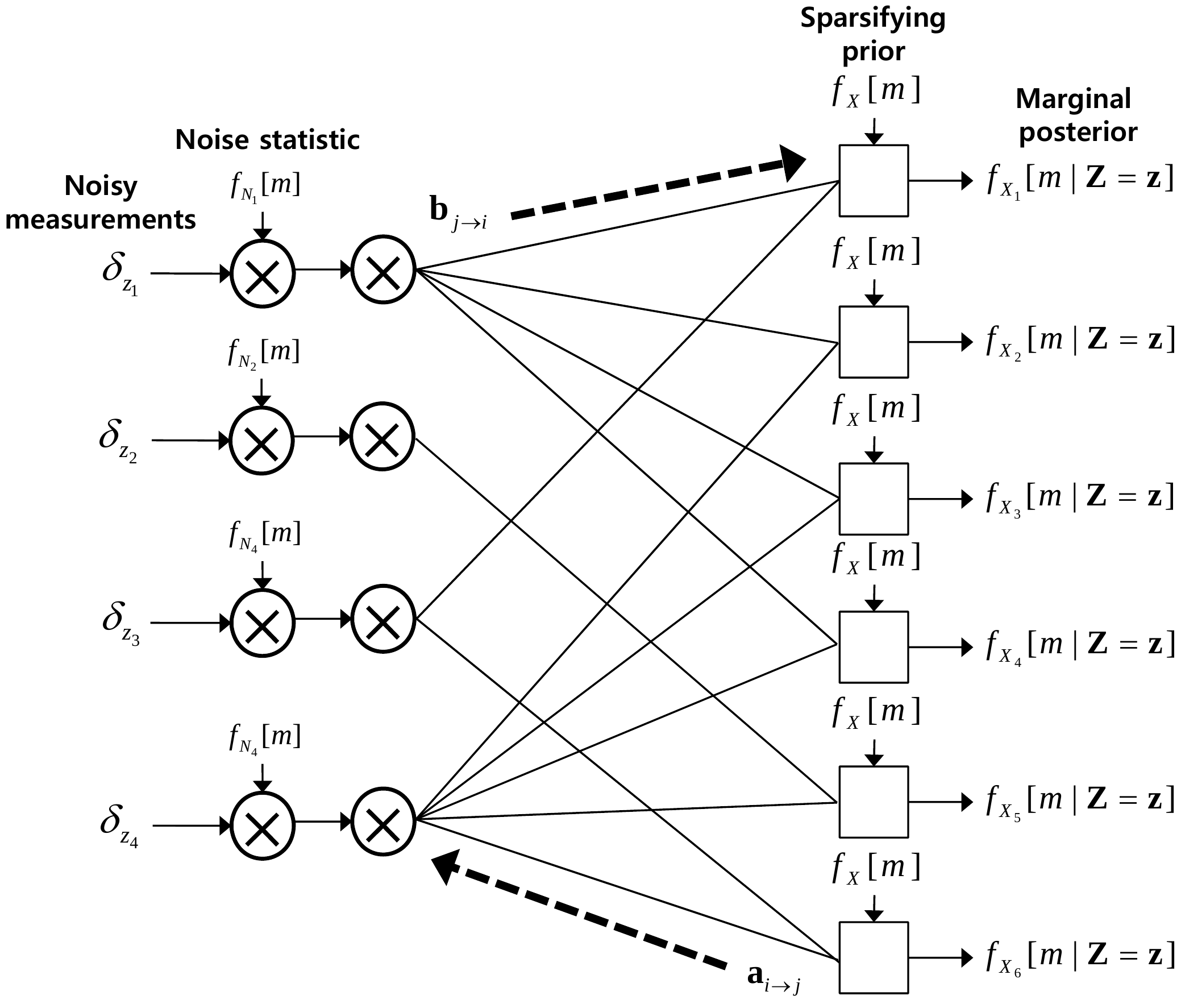}
\caption{Graph representation of the sampled-message based BP ($N=6,
M=4, L=2$), where $\otimes$ is the operator for the linear
convolution of PDFs and $\delta_{z_j}$ indicates the delta function
peaked at $z_j$.} \label{fig:Fig4-1}
\end{figure}

\subsection{Sampled-message based Belief Propagation}
BP  provides the marginal posterior for the hypothesis test in
\eqref{eq:eq3-7}  for each element. Since the signal is real-valued,
each BP-message takes the form of a PDF, and the BP-iteration
consists of density-message passing. We provide a brief summary of
density-message passing in Appendix I. To implement the BP passing
density-messages, we consider the sampled-message approach which has
been discussed by Sudderth \emph{et al.} \cite{non_para_BP}, and
Noorshams \emph{et al.} \cite{Noorshams}. For the sparse recovery,
Baron \emph{et al.} applied the approach in
\cite{CS-BP1},\cite{CS-BP2}. The main strength of the
sampled-message approach is adaptivity to various signal models. In
addition, it shows faster convergence than the parametric-message
approach \cite{Guo06},\cite{para_BP},\cite{rangan} if the sampling
step size is sufficiently fine. The reason is that the
sampled-message approach does not use any model approximation or
reduction for the message representation during the iteration. Based
on \eqref{eq:eq4-3} and \eqref{eq:th1-1} from Appendix I, we will
provide a practical message update rule of the sampled-message based
BP for the proposed algorithm.

In our implementation, we set the sampling step $T_s$ on the basis
of \emph{the three sigma-rule} \cite{sigmarule}, given as
\begin{eqnarray} \label{eq:eq4-4}
T_s=\frac{2 \cdot 3 \sigma_{X_{1}}}{N_d}
\end{eqnarray}
where $N_d$ is the number of samples for a density-message. Hence,
we define the sampling process of the PDFs as
\begin{eqnarray}
\begin{array}{l}
{\rm{Samp}}\left[ {f(x)} \right] := f(m{T_s} - 3{\sigma_{X_{1}}}) \\
\,\,\,\,\,\,\,\,\,\,\,\,\,\,\,\,\,\,\,\,\,\,\,\,\,\,\,\,\,\,\,\,\,\,=f[m]\,\,\,{\rm{
for }}\,\,\,m = 0\sim {N_d}-1,
\end{array}
\end{eqnarray}
where $\rm{Samp}[\cdot]$ denotes the sampling process. Hence, the
density-messages are treated as vectors with size $N_d$ in this
approach.

Let $\mathbf{a}_{i \rightarrow j}[m] \in [0,1]^{N_d}$ denote a
sampled density-message from $X_{i}$ to $Z_j$, called the signal
message;, and $\mathbf{b}_{j \rightarrow i}[m] \in [0,1]^{N_d}$ is
the message from $Z_j$ to $X_{i}$, called the measurement message.
The signal message $\mathbf{a}_{i \rightarrow j}[m]$ includes
information on the marginal posterior
$f_{X_i}(x|\mathbf{Z}=\mathbf{z})$, being obtained from
\eqref{eq:eq4-3} by simply replacing the measurement densities,
$f_{Z_k}(z|X_i=x_i)$,  with the measurement messages of the previous
iteration except the intrinsic information. That is,
\begin{eqnarray}\label{eq:eq4-5}
\mathbf{a}_{i \rightarrow j}^l[m]=\eta\left[ {{f_{X}}[m]  \times
\prod\limits_{k \in N_{\mathcal{V} }(i)\backslash\{j\}} {{\bf{b}}_{k
\rightarrow i}^{l-1} }[m] } \right],
\end{eqnarray}
$\text{ for all } (i,j) \in \mathcal{E}$, where  $\eta[\cdot]$ is
the normalization function to make $ \sum{\mathbf{a}_{i\rightarrow
j}[m] }=1$. Similarly, the measurement message $\mathbf{b}_{j
\rightarrow i}[m]$ includes information on  the measurement density
$f_{Z_j}(z|X_i=x_i)$, being obtained from the expression of
\eqref{eq:th1-1} by replacing the associated marginal posteriors,
$f_{X_k}(x|\mathbf{Z}=\mathbf{z})$, with the signal messages, that
is,
\begin{eqnarray}\label{eq:eq4-6}
\begin{array}{l}
\mathbf{b}_{j \rightarrow i}^l[m]\\ =\underbrace
{{f_{\mathcal{N}}}[m;{z_j},\sigma
_N^2]}_{={\rm{Samp}}[\mathcal{N}(n;z_j,\sigma_N^2)]}
 \otimes \left(\bigotimes
\limits_{k \in N_{\mathcal{C}}(j)\backslash \{i\} } {\mathbf{a}_{k
\rightarrow j}^{l}[-m]} \right),
\end{array}
\end{eqnarray}
where $\otimes$ is the operator for the linear convolution of PDFs.
In addition, under the Gaussian noise assumption,  we know
$\delta_{z_j} \otimes f_{N_j}(n)=\mathcal{N}(n;z_j,\sigma_N^2)$ in
\eqref{eq:eq4-6} based on \eqref{eq:th1-1}. Here, we note that the
measurement message calculation  utilizes the noise statistics
distinctively from that of the standard CS-BP [see (7) in
\cite{CS-BP2}]. This improvement remarkably enhances the recovery
performance in the low SNR regime even though the noise statistic
loses its effect with sufficiently high SNR.

The convolution operations in \eqref{eq:eq4-6} can be efficiently
computed by using the \emph{Fast fourier transform} (FFT).
Accordingly, we express for the measurement message calculation as
\begin{eqnarray}\label{eq:eq4-7}
\begin{array}{l}
{\bf{b}}_{j \to i}^l[m]\\= \mathcal{F}^{ - 1} \left[
\mathcal{F}\left[ {f_{\mathcal{N}}}[m;{z_j},\sigma _N^2] \right]
\left(\prod \limits_{k } {\mathcal{F}[\mathbf{a}_{k \rightarrow
j}^{l}[-m]]} \right) \right]
\end{array}
\end{eqnarray}
where $\mathcal{F}[\cdot]$ denotes the FFT operation. Therefore, for
efficient use of FFT, the sampling step $T_s$ should be
appropriately chosen such that $N_d$ is two's power. In fact, the
use of the FFT brings a small calculation gap since the FFT-based
calculation performs a circular convolution that produces output
having a heavy tail. The heaviness increases as the column weights
of $\mathbf{\Phi}$ increases. However, the difference is can be
ignored, especially when the messages are bell-shaped densities.

%\begin{figure}[!t]
%\centering
%\includegraphics[width=7cm]{fig4-2.eps}
%\caption{Calculation gap between use of linear convolution and
%FFT-based convolution in measurement message calculation.}
%\label{fig:Fig4-2}
%\end{figure}
\begin{figure}[!t]
\centering
\includegraphics[width=8cm]{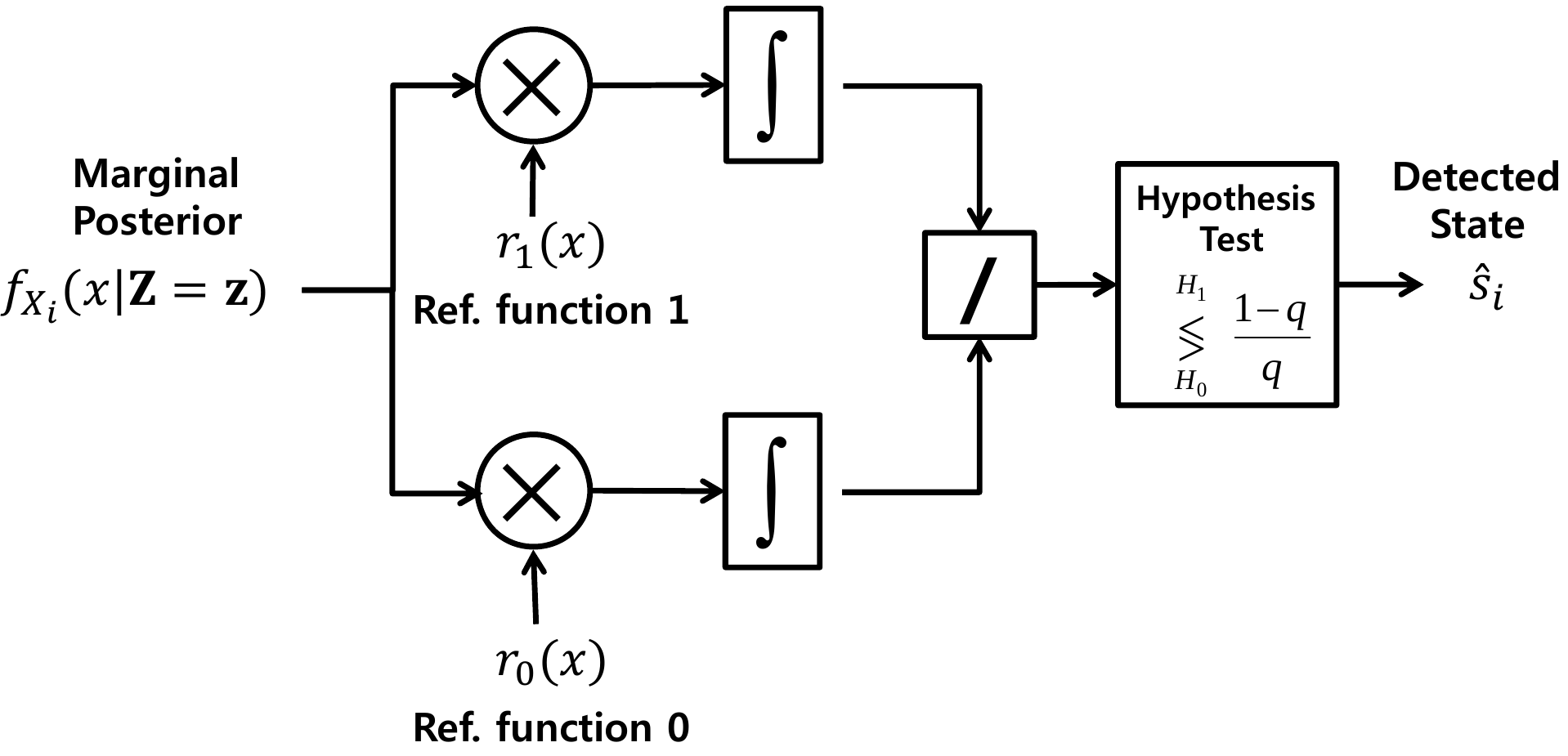}
\caption{Bayesian hypothesis test for support detection where
$r_0(x):=\frac{{f_{X}( x |S = 0) }}{{f_{X}( x )
}},r_1(x):=\frac{{f_{X}( x |S = 1) }}{{f_{X}( x ) }}$ are reference
functions consisting of the prior knowledge.} \label{fig:Fig4-3}
\end{figure}

Finally, an sampled approximation of the marginal posterior
${f_{{X_{i}}}}[m|\mathbf{Z}={\bf{z}}]  \text{ for }\,\, m=0 \sim
N_d-1$ is obtained after a certain number of iteration $l^*$ via the
update rule in \eqref{eq:eq4-5},\eqref{eq:eq4-6}, as follows:
\begin{eqnarray} \label{eq:eq4-8}
\begin{array}{l}
\text{Samp}[f_{X_{i}}(x|\mathbf{Z}=\mathbf{z})] \approx
f_{X_{i}}[m|\mathbf{Z}=\mathbf{z}]\\
\,\,\,\,\,\,\,\,\,\,\,\,\,\,\,\,\,\,\,\,\,\,\,\,\,\,\,\,\,\,\,\,\,\,\,\,\,\,\,\,\,\,\,\,\,\,\,\,\,\,\,\,
=\eta\left[
{ f_{X}[m] \times \prod\limits_{j \in N_{\mathcal{V}}(i)}
{{\mathbf{b}_{j \rightarrow i}^{l^*}[m]} } } \right].
\end{array}
\end{eqnarray}

\subsection{Bayesian Hypothesis Test for Support Detection}
In order to perform the hypothesis test in \eqref{eq:eq3-7}, the
decoder needs to calculate the probability ratio  $ \frac{{\Pr \{
S_i = 0|{\bf{Z}}\} }}{{\Pr \{ S_i = 1|{\bf{Z}}\} }} $. By
factorizing over $X_i$, the hypothesis test becomes
\begin{align}\label{eq:eq4-9}
&{\frac{{\Pr \{ S_i  = 1|{\mathbf{Z}=\bf{z}}\} }}{{\Pr \{ S_i  =
0|\mathbf{Z}={\bf{z}}\} }}}\nonumber\\
&=  {\frac{{\int  {\Pr \{ S_i = 1|{\bf{Z}}=\mathbf{z},X_{i} \}
f_{X_{i}}(x |\mathbf{Z}=\mathbf{z} ) }dx }}{{\int  {\Pr \{ S_i =
0|{\bf{Z}}=\mathbf{z},X_{i} \} f_{X_{i}}( x |\mathbf{Z}=\mathbf{z} )
}dx }}} \mathop {\mathop \gtrless \limits_{{H_0}} }\limits^{{H_1}}
1.
\end{align}
In practice, we replace the marginal posterior
$f_{X_{i}}(x|\mathbf{Z}=\mathbf{z})$ with the sampled marginal
posterior $f_{X_{i}}[m|\mathbf{Z}=\mathbf{z}]$ from \eqref{eq:eq4-8}
for discrete signal processing, which is provided in Algorithm
\ref{algo1}. However, we use notations of the continuous domain in
the description of this hypothesis process. Given $X_i=x_{i}$ from
BP, the given measurements $\mathbf{z}$ do not provide any
additional information on $S_i$; hence, $\Pr \{ S_i
|\mathbf{Z}={\bf{z}},X_{i} \} = \Pr \{S_i |X_{i} \}$ holds true.
Using such a fact and the Bayesian rule, the test in
\eqref{eq:eq4-9} is rewritten as
\begin{eqnarray}\label{eq:eq4-10}
{\frac{{\int  { r_1(x)
 f_{X_{i}}(x |\mathbf{Z}=\mathbf{z} )}dx }}{{\int {  r_0(x) f_{X_{i}}(x |\mathbf{Z}=\mathbf{z} ) }dx }}} \mathop
{\mathop \gtrless \limits_{{H_0}} }\limits^{{H_1}} {\frac{{\Pr \{ S
= 0\} }}{{\Pr \{ S= 1\} }}},
 \end{eqnarray}
where $\frac{\Pr\{S=0\}} {\Pr \{S=1\} }={\frac{(q-1)}{q}}$, and
$r_0(x),r_1(x)$ are reference functions consisting of the prior
knowledge defined as
\begin{eqnarray}\label{eq:eq4-11}
r_0(x):=\frac{{f_{X}( x |S = 0) }}{{f_{X}( x ) }},\,\,\,
r_1(x):=\frac{{f_{X}( x |S = 1) }}{{f_{X}( x ) }}.
\end{eqnarray}

The process to calculate the probability ratio in \eqref{eq:eq4-10}
is described as a block diagram in Fig.\ref{fig:Fig4-3}. This
process has a similar structure to matched filtering in
communication systems, where the detector determines supportive
state of $X_{i}$ by measuring inner products between the marginal
posterior and reference functions. In addition, we emphasize that
this BHT-based detector is only compatible with the sampled-message
based BP because the BHT-process requires full information on the
signal posterior which cannot be provided through the
parametric-message based BP.

\begin{algorithm}[!t]
\caption{BHT-BP}\label{algo1}
\begin{algorithmic}[0]
\Require Noisy measurements $\mathbf{z}$, Sensing matrix
$\mathbf{\Phi}$, Sparsfying prior $f_{X}(x)$, Noise statistic
$f_{N}(n)$, Sampling step $T_s$, Number of the BP-iterations $l^*$.

\Ensure Reconstructed signal
$\widehat{\mathbf{x}}_{0,\text{BHT-BP}}$, Detected state vector
$\widehat{\mathbf{s}}$.

\State {\bf{1) Sampled-message based BP:}} \State set $\mathbf{b}_{j
\rightarrow i}^{l=0}=\mathbf{1}\text{ for all } (i,j) \in
\mathcal{E}$ \For{$l=1$ \textbf{to} $l^*$} \For{$i=1$ \textbf{to}
$N$} \State set $\mathbf{a}_{i \rightarrow j}^l[m]=\eta\left[
{{f_{X}}[m]  \times \prod\limits_{k \in N_{\mathcal{V}
}(i)\backslash\{j\}} {{\bf{b}}_{k \rightarrow i}^{l-1} }[m] }
\right]$ \EndFor

\For{$j=1$ \textbf{to} $M$} \State $\begin{array}{l}\text{set
}\mathbf{b}_{j \rightarrow i}^l[m]
\\=f_{\mathcal{N}}[m;{z_j},\sigma _N^2]
 \otimes \left(\bigotimes
\limits_{k \in N_{\mathcal{C}}(j)\backslash \{i\} } {\mathbf{a}_{k
\rightarrow j}^{l}[-m]} \right)\end{array}$ \EndFor \EndFor

\For{$i=1$ \textbf{to} $N$} \State set
$f_{X_{i}}[m|\mathbf{Z}=\mathbf{z}]=\eta\left[ { f_{X}[m] \times
\prod\limits_{j \in N_{\mathcal{V}}(i)} {{\mathbf{b}_{j \rightarrow
i}^{l^*}[m]} } } \right]$ \EndFor

\State {\bf{2) BHT for Support Detection:}}
 \State set $\gamma= q/(1-q) $
    \For{$i=1$ \textbf{to} $N$}

        \If {${\frac{{\sum_{m}  {   r_1[m]f_{X_{i}}[m
|\mathbf{Z}=\mathbf{z} ]} }}{{\sum_{m} {  r_0[m]f_{X_{i}}[m
|\mathbf{Z}=\mathbf{z} ] } }}}
>  \frac{(1-q)}{q}$} set $\widehat{s}_i =1$
        \Else  {} set $\widehat{s}_i =0$
        \EndIf

    \EndFor
    \State {\bf{3) Signal Value Estimation:}}
\State set ${\widehat {\bf{x}}_{0,\text{BHT-BP}}} = {\bf{E}}\left[
{{{\bf{X}}}|{\bf{S}} = \widehat{{\bf{s}}},\mathbf{Z}={\bf{z}}}
\right]$
\end{algorithmic}
\end{algorithm}

\subsection{Computational Complexity}
In the sampled-message approach,  the density-messages in BP are
vectors with size $N_d$. Therefore, the decoder requires $O(LN_d)$
flops  to calculate a signal message $\mathbf{a}_{i \rightarrow j}$
and $O( {\frac{{NLN_d }}{M}\log N_d })$ flops for a measurement
message $\mathbf{b}_{j \rightarrow i}$ per iteration, where $L$
denotes the column weight of the sparse sensing matrix
$\mathbf{\Phi}$. In addition, the cost of the FFT-based convolution
is $O(N_d\log N_d)$ if we assume the row weight is $NL/M$ using
average sense. Hence, the per-iteration cost of the BP-process is
$O(NLN_d+M{\frac{{NLN_d }}{M}\log N_d })\approx O(NLN_d \log N_d)$
flops. For the hypothesis test, the decoder requires $O(N_d)$ flops
to generate the probability ratio. The cost for the hypothesis test
is much smaller than that of BP; therefore, it is ignored. For the
MMSE estimator to find signal values, the cost can be reduced upto
$O(KM)$ flops by applying QR decomposition \cite{Bjorck}. Thus, the
total complexity of the proposed algorithm is $O\left( l^* \times
NLN_d \log N_d + KM \right)$ flops and it is further simplified to
$O(l^* \times N+KM )$ since $L$ and $N_d$ are fixed. In addition, it
is known that the recovery convergence via BP is achieved with
$O(\log N)$ iterations \cite{CS-BP2},\cite{Mackey}. Therefore, we
finally obtain $O(N \log N +KM)$ for the complexity of BHT-BP. We
note here that the complexity of BHT-BP is much relaxed from that of
the naive MMSE sparse estimator $O(2^N)$, given in \eqref{eq:eq3-5}.

\section{Exemplary Discussion for Proposed Algorithm}
 One can argue the DD-structure of the proposed algorithm is an abuse since the
marginal posteriors from the BP already provides full information on
the signal. Namely, this means that
$f_{X_{i}}[m|\mathbf{Z}=\mathbf{z}]$ in \eqref{eq:eq4-8} contains
the perfect knowledge for detection and estimation of  $x_{0,i}$.
Yes it is true, but our claim is that the MAP-based algorithms
\cite{CS-BP1},\cite{CS-BP2} which solve the problem in
\eqref{eq:eq_3} are not utilizing the full knowledge of the marginal
posterior.

 In this section, we  show two weakpoints of the MAP-approach
which finds each scalar estimate $\widehat{x}_{0,i}$ only using the
peak location of the marginal posterior, through examples. We then
discuss how the proposed algorithm can remedy such problematic
behavior of the MAP, and utilize the posterior knowledge more
efficiently. We claim that the proposed algorithm has strength in
two aspects as given below:
\begin{enumerate}
\item \emph{Robust support detection against additive noise}: The
BHT-detector in the proposed algorithm correctly catches the
supportive state given $i \in supp(\mathbf{x}_0)$ even under severe
additive noise.
\item \emph{Removing quantization effect caused by the sampled-message based
BP}: The quantization effect degrades performance of the MAP-based
algorithms in both the signal value estimation and the support
detection. In the proposed algorithm, the DD-structure  removes the
quantization effect from the signal value estimation, and the
BHT-detector reduces the chance of misdetection of the supportive
state given $i \notin supp(\mathbf{x}_0)$ by applying the knowledge
of $x_{min}$ to the reference functions.
\end{enumerate}

\begin{figure}
\centering
\includegraphics[width=8cm]{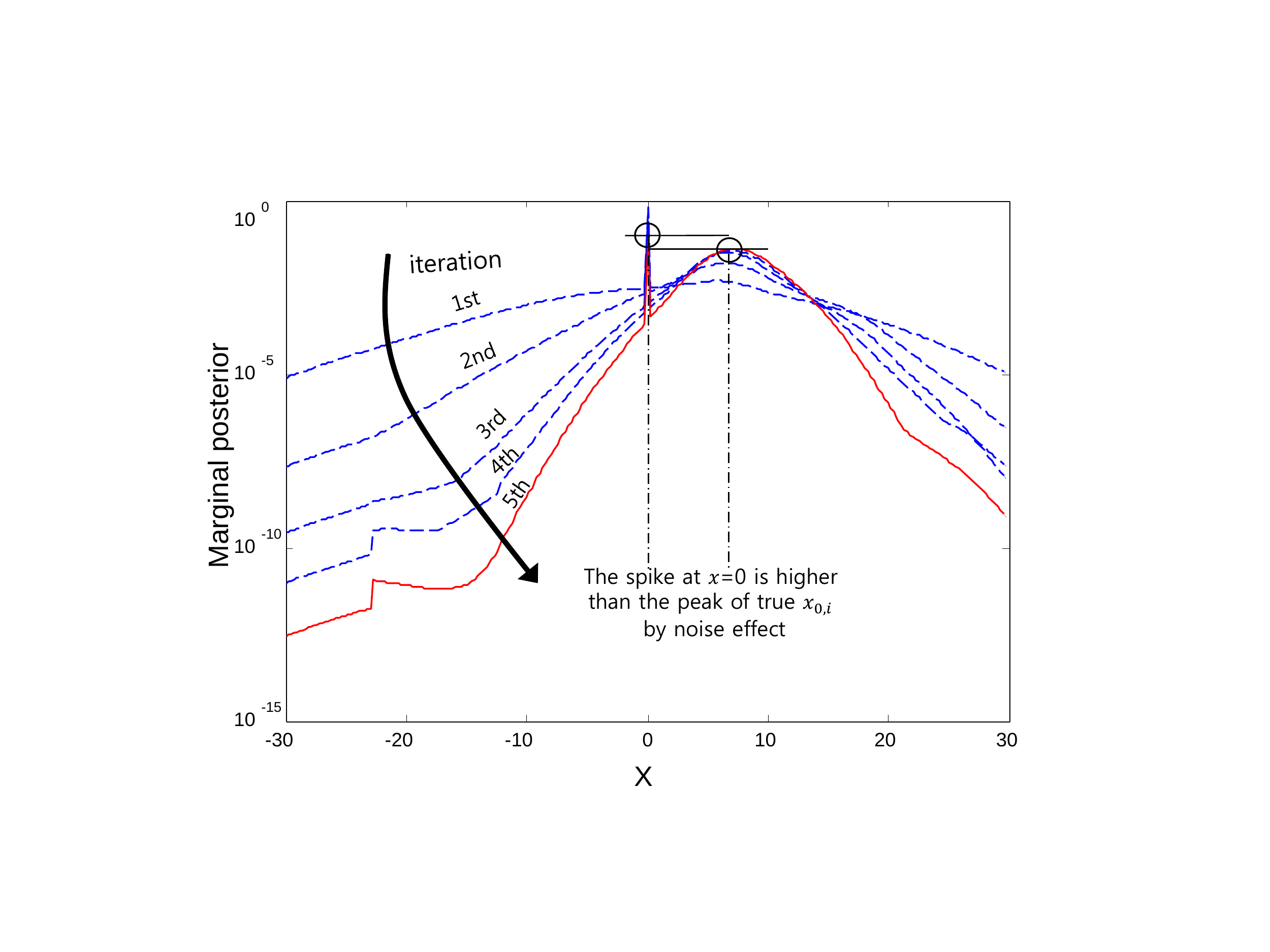}
\caption{Example of  a marginal  posterior
$f_{X_{i}}(x|\mathbf{Z}=\mathbf{z})$ in BP-iteration under severe
noise (SNR=10 dB), where $x_{0,i}$ originally has nonzero
($x_{0,i}=6.7)$, the minimum value is $x_{min}=1.25$. }
\label{fig:Fig4-5}
\bigskip
\centering
\includegraphics[width=8cm]{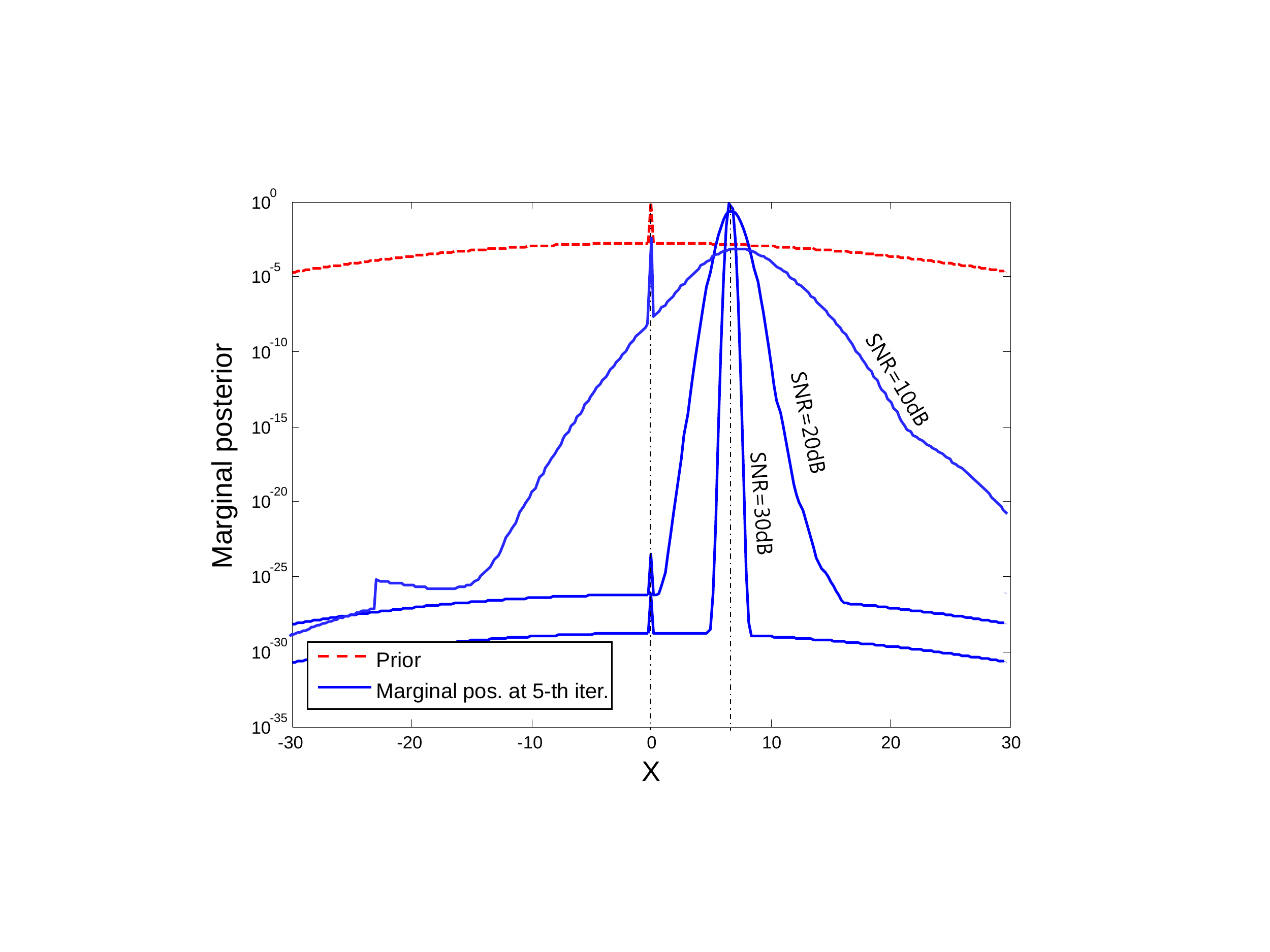}
\caption{Example of  a marginal  posterior
$f_{X_{i}}(x|\mathbf{Z}=\mathbf{z})$ over SNR where $x_{0,i}$
originally has nonzero ($x_{0,i}=6.7)$, the minimum value is
$x_{min}=1.25$, and we use 5 BP-iterations to approximate
$f_{X_{i}}(x|\mathbf{Z}=\mathbf{z})$.} \label{fig:Fig4-5-2}
\end{figure}

\subsection{Robust Support Detection against Additive Noise}
The additive noise spreads probability mass in the marginal
posterior $f_{X_{i}}(x|\mathbf{Z}=\mathbf{z})$, leading to
difficulty in the supportive state detection  via the MAP-approach
given $i \in supp(\mathbf{x}_0)$. For example, Fig.\ref{fig:Fig4-5}
shows marginal posterior obtained from severe noisy measurements
(SNR=10 dB) via BP-iteration, where the signal value is originally
$x_{i,0}=6.7$, hence $i \in supp(\mathbf{x}_0)$. We note that the
posterior is approximately composed of a zero-spike and a
slab-density with  certain probability weights. In
Fig.\ref{fig:Fig4-5}, it is shown that the center of the
slab-density is moving to the true value $x=6.7$ over the
iterations, but after 5 iterations the mass at $x=6.7$, \emph{i.e.},
$f_{X_{i}}(x=6.7|\mathbf{Z}=\mathbf{z})$, is still smaller than that
of the zero-spike. The reason is that the noise effect spreads the
slab-density over near values, making the peak at $x=6.7$ to be
blunt. We call this spreading effect as \emph{density spread}.

Fig.\ref{fig:Fig4-5-2} more clearly describes the density spread in
the marginal posterior according different SNR levels. When SNR
level is sufficiently high (see cases more than SNR=20 dB in
Fig.\ref{fig:Fig4-5-2}), the MAP-approach can successfully detect
the supportive state since probability mass is concentrated on the
center of the slab-density such that
$f_{X_{i}}(x=6.7|\mathbf{Z}=\mathbf{z}) \geq
f_{X_{i}}(x=0|\mathbf{Z}=\mathbf{z})$. However, when SNR is low (see
the line of SNR=10 dB in Fig.\ref{fig:Fig4-5-2}), the MAP-approach
will fail in the detection because the zero-spike
$f_{X_{i}}(x=0|\mathbf{Z}=\mathbf{z})$ becomes the highest peak in
the marginal posterior such that
$f_{X_{i}}(x=6.7|\mathbf{Z}=\mathbf{z}) <
f_{X_{i}}(x=0|\mathbf{Z}=\mathbf{z})$ due to the density spread. The
density spread does not cause errors given $i \notin
supp(\mathbf{x}_0)$ since in this case the center of the
slab-density stays at $x=0$ during the BP-iteration, regardless of
the noise level.

In contrast, the BHT-detector in the proposed algorithm, decides the
supportive state by considering the density spread effect. In
\eqref{eq:eq4-10}, the inner products between $r_1(x),r_0(x)$ and
$f_{X_{i}}(x|\mathbf{Z}=\mathbf{z})$ measure portions of the
marginal posterior corresponding to $i \in supp(\mathbf{x}_0)$ and
$i \notin supp(\mathbf{x}_0)$ respectively. Since the inner products
are associated with the entire range of the $x$-axis rather than
specific point-mass, the BHT-detector can decide the supportive
state by incorporating all spread mass due to noise. Therefore, the
BHT-detector can success in the detection even in the SNR=10 dB case
of Fig.\ref{fig:Fig4-5-2}. This example supports that BHT-detector
has ability to detect the signal support more robustly against noise
than the MAP-approach.

\begin{figure}
\centering
\includegraphics[width=8cm]{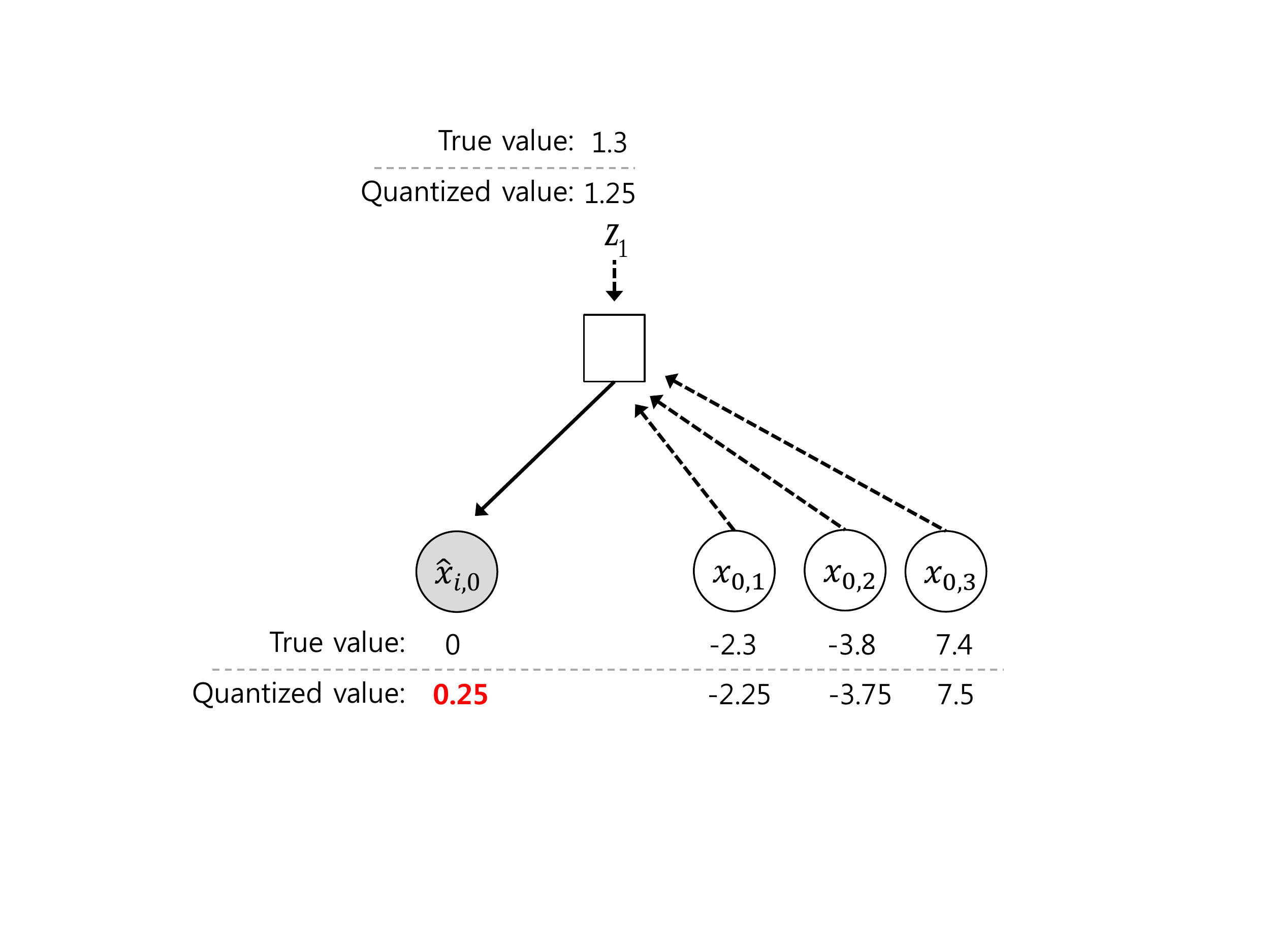}
\caption{Example of measurement message passing under quantization
caused by the sampled-message based BP where $T_s=0.25$ }
\label{fig:Fig4-6-3}
\end{figure}

\subsection{Removing Quantization Effect caused by the
Sampled-message Based BP} \subsubsection {In signal value
estimation}
 Under the use of the sampled-message based BP,
quantization error is inevitable in the signal value estimation.
When we apply rounding for the quantization, the MSE error is
uniformly distributed with zero-mean and variance according to the
sampling step size $T_s$, given as
\begin{align}  \label{eq:eq4-12}
{\bf{E}}\left\| {{Q_{{T_S}}}[{X_{i \in {\rm{supp}}({\bf{X}})}}] -
{X_{i \in {\rm{supp}}({\bf{X}})}}} \right\|_2^2
=\frac{T_s^2}{12}\nonumber\\=\left(\frac{2 \cdot 3
\sigma_{X_{1}}}{N_d}\right)^2/12,
\end{align}
where ${Q_{{T_s}}}\left[  \cdot  \right]$ denotes the quantization
function with $T_s$. Therefore,  MSE performance of the MAP-based
algorithms with the sampling cannot exceed the limit given by
\eqref{eq:eq4-12} even if the level of additive noise is extremely
low. To relax this limit, we can increase memory size $N_d$ or
confine the range of signal value by decreasing $\sigma_{X_{1}}$.
However, such methods are impractical and restrict flexibility of
signal models in the recovery. The DD-structure, described in
\eqref{eq:eq3-6} and \eqref{eq:eq3-6-2}, removes this weakpoint
since the signal values are evaluated using an MMSE estimator in
\eqref{eq:eq3-6-1} independently of $N_d$ once the detected support
is given. Furthermore, the probability ratio for the BHT-detection
 can be generated from sampled marginal
densities $f_{X_{i}}[m|\mathbf{Z}=\mathbf{z}]$ with small $N_d$.

\begin{figure}
\centering
\includegraphics[width=8cm]{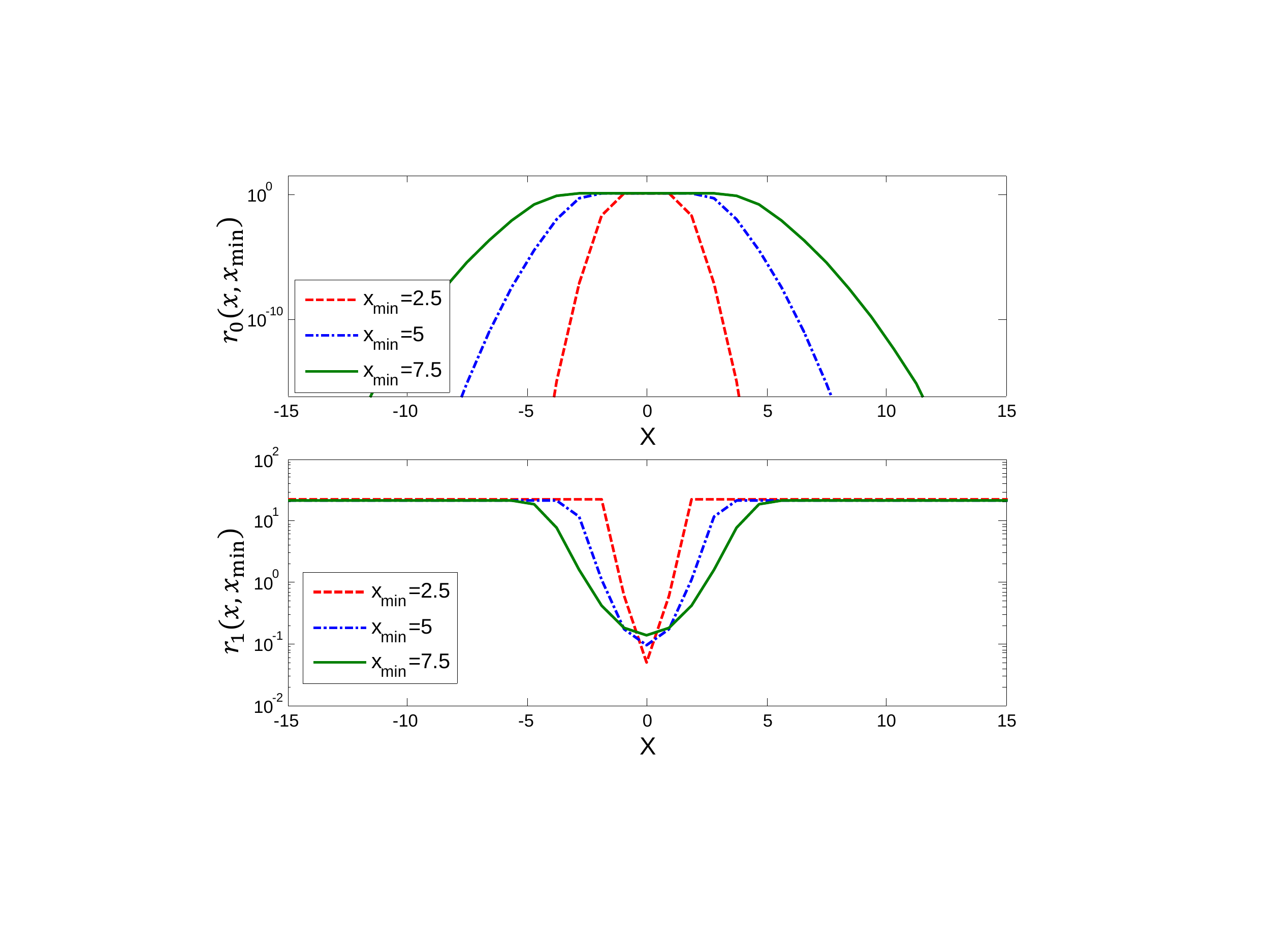}
\caption{Calibrated reference functions for Bayesian hypothesis test
according to $x_{min}$ where $c=1/6$.} \label{fig:fig4-4}
\end{figure}

\subsubsection {In  support detection}
 The quantization effect
also can cause detection error of the supportive state given $i
\notin supp(\mathbf{x}_0)$, limiting performance of the MAP-approach
in the high SNR regime. In order to explain such  cases, we provide
an example of the measurement message-passing with respect to
$x_{0,i}=z_1-(x_{0,1}+x_{0,2}+x_{0,3})$ as shown in
Fig.\ref{fig:Fig4-6-3}. In this example, we try to find the value of
$\widehat{x}_{0,i}$ given $x_{0,1}=-2.3$, $x_{0,2}=-3.8$,
$x_{0,3}=7.4$, $z_1=1.3$, and $i \notin supp(\mathbf{x}_0)$, under
noiseless setup. Note that in practical operation of the algorithm,
the value of each element corresponds to the peak location in the
$x$-axis of each density-message. In the sampled-message based BP,
the  message sampling causes quantization of the values such that we
have $Q_{T_s}[x_{0,1}]=-2.25$, $Q_{T_s}[x_{0,2}]=-3.75$,
$Q_{T_s}[x_{0,3}]=7.5$, $Q_{T_s}[z_1]=1.25$ with the step size
$T_s=0.25$ where $Q_{T_s}[\cdot]$ denotes the quantization function.
In this case, we can simply infer that this measurement message
passes a corrupted value $\widehat{x}_{0,i}=0.25$ which is not
matched with $i \notin supp(\mathbf{x}_0)$. If most of the
measurement messages to ${x}_{0,i}$ has such corruption, the
marginal posterior of ${x}_{0,i}$ will have the peak at an erroneous
location, leading to detection error based on the  MAP-approach. The
same event can occur given $i \in supp(\mathbf{x}_0)$. However, the
effect is not significant because the  case of $i \in
supp(\mathbf{x}_0)$ does not bring about the support detection
error. In addition, this type of errors is remarkable particularly
in the high SNR regime since such corruption is covered by additive
noise when SNR is low. Accordingly, the support detection
performance of the MAP-approach shows an error-floor in the high SNR
regime.

 Here, we aim to show how the proposed BHT-detector utilizes the prior
knowledge of $x_{min}$  to remove the error-floor of the MAP
approach. As we mentioned in Section III-A, we consider the signal
model which has $x_{min}$ as a key parameter, according to the
result of Wainright \emph{et al.}
\cite{Wainwright1},\cite{Wainwright2}. The Wainright' result
revealed that the regulation of $x_{min}$ is imperative for perfect
support recovery under noisy setup. This means that we can have
additional prior knowledge to the sparse recovery. From
\eqref{eq:eq3-2}, the knowledge of $x_{min}$ tells us that there
exist no nonzero elements which have value within $ |x_{0,i}| <
x_{min}$. The BHT-detector reflects $x_{min}$ to calibrate the
reference functions. Rather than the functions given in
\eqref{eq:eq4-11}, we use its calibrated version given as
\begin{eqnarray} \label{eq:eq4-13}
r_0(x,x_{min})&:=&\frac{{f_{X}( x |S = 0;x_{min}) }}{{f_{X}( x
;x_{min}) }},\nonumber\\
r_1(x,x_{min})&:=&\frac{{f_{X}( x |S = 1) }}{{f_{X}( x;x_{min} ) }},
\end{eqnarray}
to improve the performance for the Gaussian signal model, where
\begin{eqnarray}
\begin{array}{l}
f_{X}( x |S = 0;x_{min}) = \mathcal{N}(x;0,c\,x_{min}),\\
f_{X}(x;x_{min})= qf_{X}( x |S = 1)+ (1 - q)f_{X}( x |S =
0;x_{min}),
\end{array}\nonumber
\end{eqnarray}
with a constant $c \in \mathbb{R}$. For signed signals, we simply
use $x_{min}=c\cdot \sigma_{X_{1}}/2$.

This calibration depresses the weight of $r_1(x,x_{min})$ and puts
more weight to $r_0(x,x_{min})$ over $ |x_{0,i}| < x_{min}$, as
shown in Fig.\ref{fig:fig4-4}. This implies that the detector
excludes elements within $ |x_{0,i}| < x_{min}$ from the signal
support set. Therefore, we can eliminate the misdetection cases
given $i \notin supp(\mathbf{x}_0)$ such as the example in
Fig.\ref{fig:Fig4-6-3}, removing the error-floor in the high SNR
regime effectively.

\begin{table}[!t]
\renewcommand{\arraystretch}{1.3}
 %if using array.sty, it might be a good idea to tweak the value of
 %\extrarowheight as needed to properly center the text within the cells
\caption{List of sparse recovery algorithms in the experiment}
\label{table2}
 \centering
\begin{tabular}{||c||c|c|c||}
\hline\hline
Algorithm & Complexity        & Type of $\Phi$    & Use of no-    \\
          & for recovery      &     &              ise  statistic               \\
\hline \hline
BHT-BP    & $O(N\log N+KM)$        & sparse-binary              & Yes                \\
 \hline
Standard   & $O(N\log N)$                  & sparse-binary   & No             \\
 CS-BP   &                  &    &              \\
\hline
CS-BP-NS   & $O(N\log N)$                  & sparse-binary        & Yes          \\
\hline
SuPrEM    & $O(N\log N)$               & sparse-binary          & Yes        \\
\hline
BCS     & $O(NK^2)$                   & dense-Gaussian         & No            \\
\hline
L1-DS  & $\Omega(N^3)$           & dense-Gaussian              & No        \\
\hline\hline
\end{tabular}
\end{table}

\section{Experimental Validation} \label{Numresult}
We support our investigation for the proposed algorithm via
experimental validation. We performed two types of the experiments
for this validation as given below:
\begin{enumerate}
\item \emph{SER performance over SNR}: To support
our claims for support detection based on BHT, we simulate the SER
performance, described in \eqref{eq:eq3-8}, as a function of SNR,
compared to that of the MAP-approach used in CS-BPs,
\item \emph{MSE comparison over SNR}: To compare the
recovery performance of the recent NSR algorithms listed in
Table.\ref{table2} and the oracle estimator, in the presence of
noise, we examine MSE performance, described in \eqref{eq:eq3-9}, as
a function of SNR.
\end{enumerate}
Here, for SuPrEM, BCS, and L1-DS, we obtained the source codes from
each author's webpage. For CS-BP, we implemented it using the
sampled-message based approach and the spike-and-slab prior, making
the algorithm into two different versions: 1) the standard CS-BP
which is the original by the corresponding papers
\cite{CS-BP1},\cite{CS-BP2}, and 2) CS-BP-NS which utilizes the
noise statistic in the BP-iteration. In implementation of the
sampled-message based BP for CS-BPs and BHT-BP, we used $N_d=256$
such that the sampling step for the density-messages is $T_s \approx
0.1172$ with \eqref{eq:eq4-4}. For the choice of the sensing matrix,
BHT-BP and CS-BPs were sparse-binary matrices with $L=4$ such that
the matrix sparsity is $7.8\%$, as discussed in Section III-A, .
L1-DS and BCS are performed with the standard Gaussian matrix having
the equal column energy as the sparse-binary matrix, for fairness,
\emph{i.e.}, $\left\| {\mathbf{\phi}
_{j,Gaussian}}\right\|_2^2=\left\|{\mathbf{\phi}_{j,Sparse}}\right\|_2^2=L$.
In addition, SuPrEM worked with a sensing matrix generated from a
low-density frame \cite{SuPrEM}.

In these experiments, we examined the two types of signal models
defined in Section III-B, Gaussian signals and signed signals, with
$N=1024$, $q=0.05$, and $\sigma_{X_1}=5$. For the case of Gaussian
signals, we restricted the magnitude level of the signal elements to
$x_{min} \leq |x_{0,i}| \leq 3\sigma_{X_1}$ where
$x_{min}=\sigma_{X_1}/4$. In addition, we fixed the undersampling
ratio to $M/N=0.5$ because the focus of this paper is to investigate
the behavior of NSR over SNR. Also, we used Monte Carlo method with
200 trials to evaluate the performance in average sense.

\begin{figure}[!t]
\centering
\includegraphics[width=8cm]{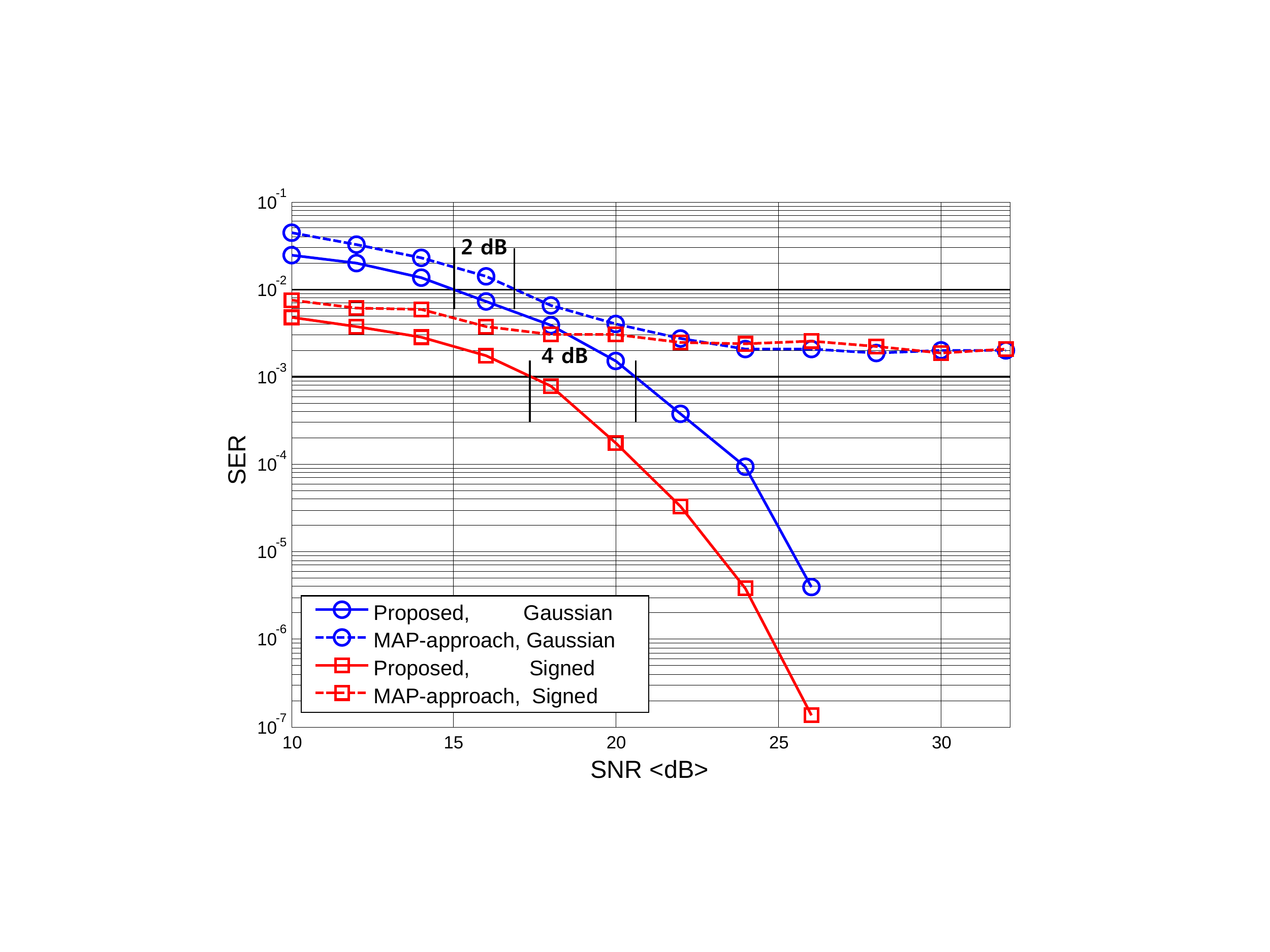}
\caption{SER performance for support detection over SNR for
$N=1024$, $q=0.05$, $M/N=0.5$, $\sigma_{X_1}=5$, $x_{min}=1.25$, and
$N_d=256\, (T_s\approx 0.1172)$.}\label{fig:Fig5-1}
\end{figure}

\subsection{SER Performance over SNR}
 Fig.\ref{fig:Fig5-1} compares a SER performance between the
 proposed algorithm and the MAP-approach used in  CS-BPs,
 where the BP-process embedded in the both algorithms utilizes the noise
statistic. We simulated the SER performance as a function of SNR for
the both signal model.

\subsubsection{Result at low SNR} In the low SNR regime, the
noise-robust property of the proposed  BHT-detector provides 2 dB
gain at SER = $10^{-2}$  from the  the MAP-approach for the Gaussian
model, as we discussed in Section V-A. In addition, we note that the
case of signed signals shows better performance than the case of
Gaussian signals with 4 dB gap at SER = $10^{-3}$. The reason is
that in the signed case the nonzero values is fixed to $X_{i\in
supp(\mathbf{X})}=\sigma_{X_1}$; therefore, the sparse patten of the
signal are rarely corrupted by noise and can be detected with less
difficulty than the Gaussian case which can have nonzero elements
$x_{0,i}<\sigma_{X_1}$.

\subsubsection{Result at  high SNR} As SNR increases, the SER
performance of the proposed algorithm shows a waterfall behavior
whereas that of  the MAP-approach shows an error-floor limited to
SER of $2 \times 10^{-3}$, for both signal models. In the MAP cases,
more SNR cannot help when the curve manifests such an irreducible
SER since the level is saturated by quantization effect caused by
the sampled-message based BP, as discussed in Section V-B. In
contrast, the SER curve of the proposed algorithm declines with more
SNR by overcoming quantization effect, removing the error-floor  of
the MAP-approach for both signal models. Therefore, this result
indicates that the proposed BHT-detector can provide the perfect
support knowledge to the signal value estimator if sufficient SNR
level is present.

\begin{figure}
\centering
\includegraphics[width=8cm]{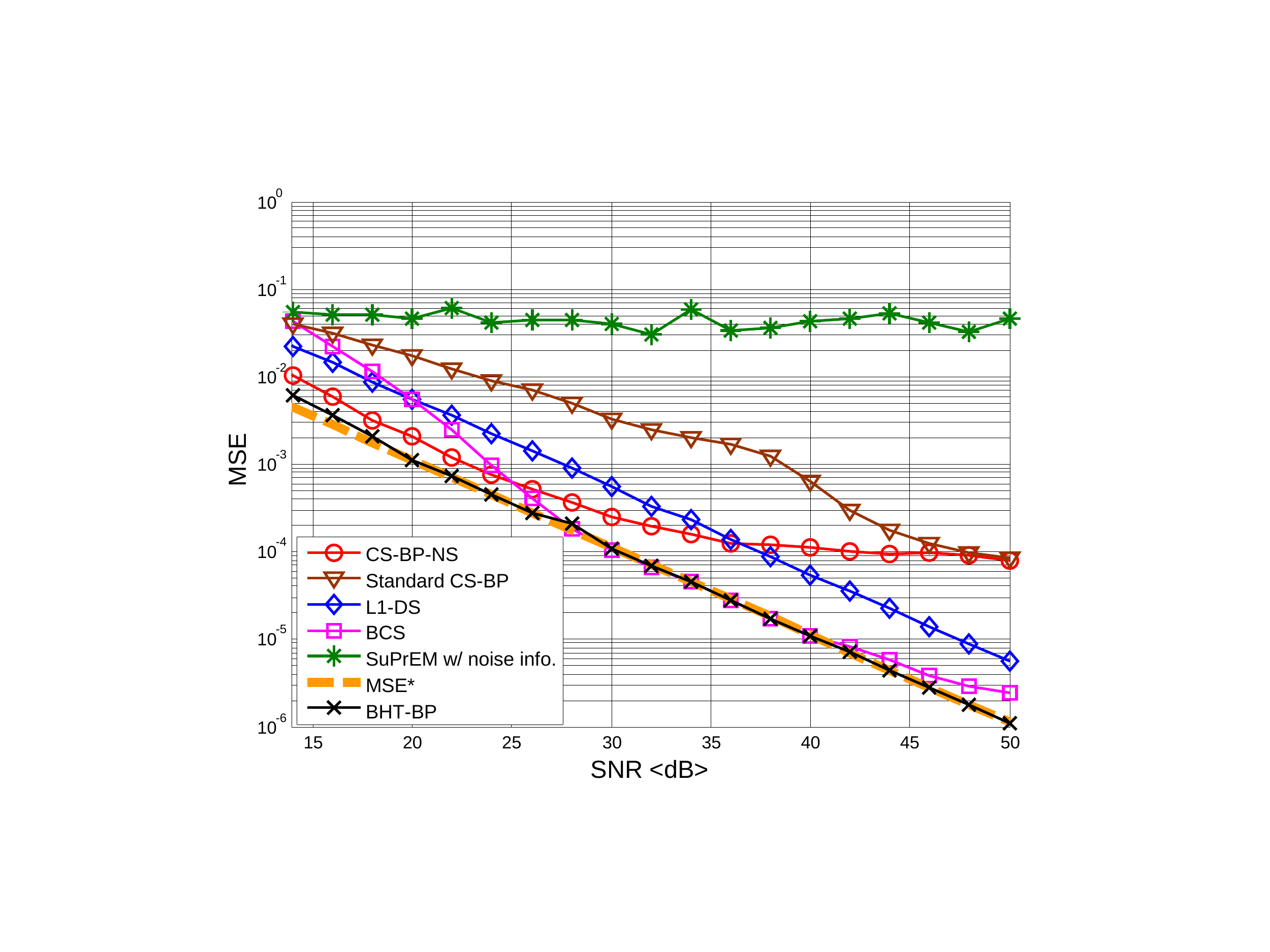}
\caption{MSE comparison over SNR for $N=1024$, $q=0.05$, $M/N=0.5$,
$\sigma_{X_1}=5$, and $N_d=256\, (T_s\approx 0.1172)$: Signed
signals case.} \label{fig:fig5-2-1}
\bigskip
\centering
\includegraphics[width=8cm]{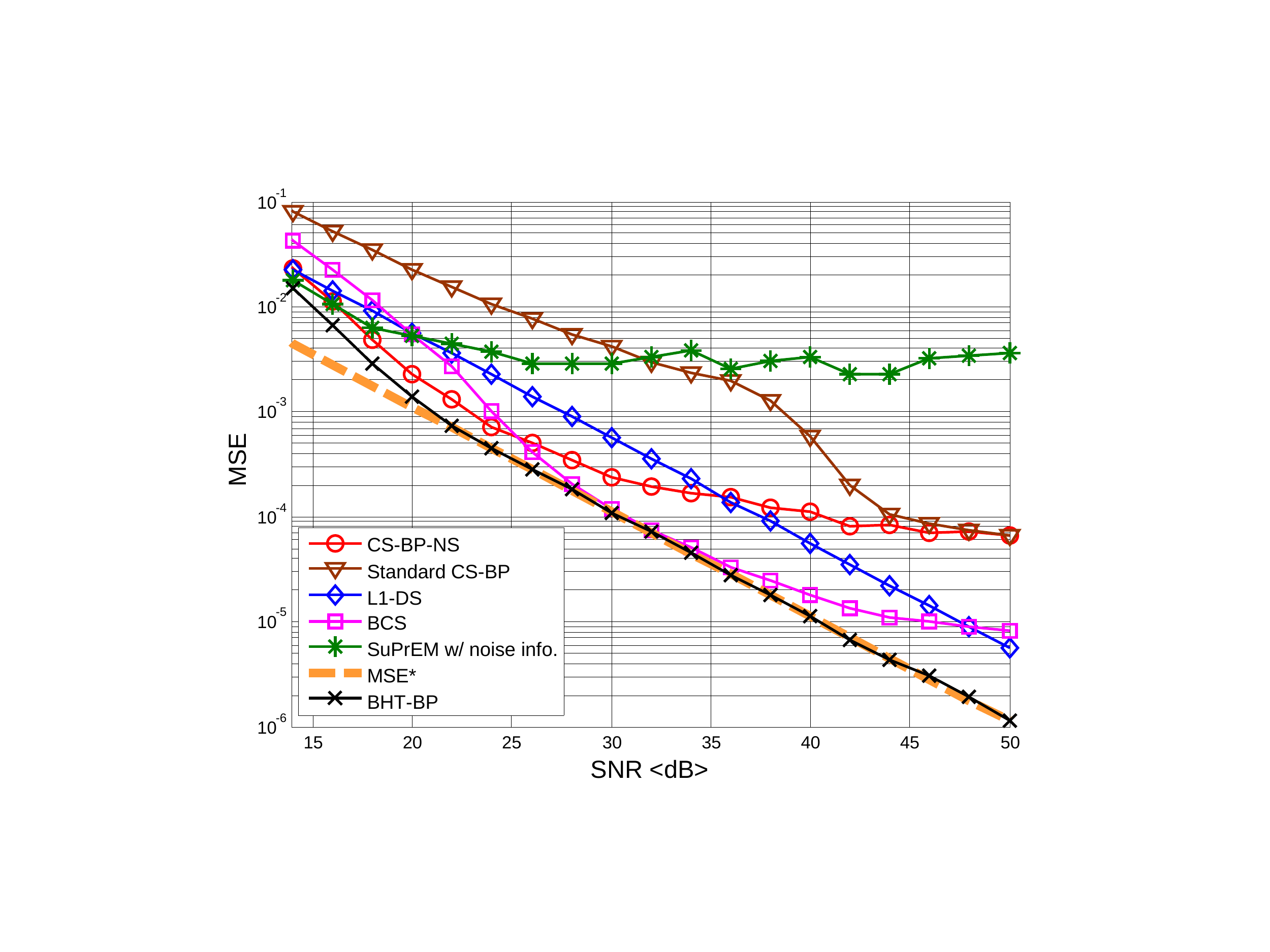}
\caption{MSE comparison over SNR for $N=1024$, $q=0.05$, $M/N=0.5$,
$\sigma_{X_1}=5$, $x_{min}=1.25$, and $N_d=256\, (T_s\approx
0.1172)$: Gaussian signals case.} \label{fig:fig5-2-2}
\end{figure}

\subsection{MSE Comparison over SNR}
 This section provides a MSE
comparison among the algorithms  in Table.\ref{table2} and the
oracle estimator, over SNR, where MSE$^*$ denotes
 the performance of the oracle estimator, given as
\begin{eqnarray} \label{eq:eq6-2}
{\text{MSE}}^* := \frac{{{\rm{Tr}} \left[ { \left(\frac{1}{{\sigma
_{X_1}^2}}{\bf{I}} + \frac{1}{{\sigma _N^2}}{{\bf{\Phi
}}_{\mathbf{s}}^T {\bf{\Phi }}_{\mathbf{s}} } \right)^{ - 1} }
\right]}}{{\left\| {{\bf{x}}_{{{0,\mathbf{s}}}} } \right\|_2^2 }}.
\end{eqnarray}
 Fig.\ref{fig:fig5-2-1} and Fig.\ref{fig:fig5-2-2} display the results for the signed signal
 and the Gaussian signal case respectively.
\subsubsection{Result at low SNR}
When the measurements are heavily contaminated by noise (below SNR
=20 dB), performance of all recovery algorithms is basically
degraded. Under such degradation, BHT-BP and CS-BP-NS outperform the
others because they are fully employing noise statistic during the
recovery process, where difference of BHT-BP and CS-BP-NS (2 dB SNR
gap at MSE = $10^{-3}$) is from types of the support detectors as we
validated in Section V-A.

The use of noise statistic remarkably affects the performance of
BP-based algorithms as discussed in Section IV-A. As a support, the
standard CS-BP shows $8 \sim 10$ dB SNR loss from CS-BP-NS for the
both signal models. For SuPrEM, even if it also includes noise
variance as a input parameter, the performance is underperformed
since SuPrEM was devised mainly for the $K$-sparse signals which
have the support set with fixed cardinality.

As SNR increases, BHT-BP  approaches the oracle performance MSE$^*$
where the case of signed signal in Fig.\ref{fig:fig5-2-1} shows
faster approaching than the case of Gaussian signals in
Fig.\ref{fig:fig5-2-2}  with approximately 4 dB gap. we note that
this gap according to the signal models is originated from the gap
in the SER performance.

\subsubsection{Result at high SNR}
In the high SNR regime, performance of the algorithms are generally
improved except SuPrEM, for the both signal models. Among the
algorithms, BHT-BP shows the most closely approaching performance to
MSE$^*$. This is because the BHT-detector provides the perfect
support knowledge beyond a certain SNR level. Although BCS shows a
competitive performance within a certain range of SNR (SNR = 28
$\sim$ 40 dB for the signed case, SNR = 30 $\sim$ 34 dB for the
Gaussian case), its performance is saturated to a certain level as
SNR becomes higher.

For CS-BPs, the use of the noise statistic in the BP-process is no
longer be effective beyond a certain SNR level. Indeed, the MSE in
Fig.\ref{fig:fig5-2-1} and Fig.\ref{fig:fig5-2-2} commonly shows
that the performance of the standard CS-BP converges to that of
CS-BP-NS beyond approximately SNR = 45 dB. In addition, after the
convergence the performances are saturated to a certain level even
with higher SNR. The cause of this saturation is the quantization
effect, as discussed in Section V-B. Using \eqref{eq:eq4-12}, we can
calculate the normalized MSE degradation by the quantization under
this experimental setup, given as
\begin{eqnarray} \label{eq:eq6-2}
\frac{{T_s^2/12}}{{{\bf{E}}\left\| {{X_{i \in
{\rm{supp}}({\bf{X}})}}} \right\|_2^2}}= 4.5786 \times {10^{ - 5}},
\end{eqnarray}
 The result in \eqref{eq:eq6-2}  closely lower bounds the MSE of
CS-BPs at SNR = 50 dB where the signed  and Gaussian cases show MSE
= $7.973 \times 10^{-5}$ and  MSE = $6.554 \times 10^{-5}$
respectively. These results exactly explain the performance loss of
CS-BPs by quantization effect, standing out BHT-BP to remove the
quantization effect using the DD-structure.

\section{Conclusion}
 The theoretical and empirical research in this paper
demonstrated that BHT-BP is a powerful algorithm for NSR from a
denoising standpoint.  In BHT-BP, we employed the DD-structure,
which consists of support detection and  signal value estimation.
The support detector is designed by a combination of the
sampled-message based BP and Bayesian hypothesis test (BHT), and the
signal value estimation performs in MMSE sense.

We have shown that BHT-BP utilizes the posterior knowledge more
efficiently  than the MAP-based algorithms, over the entire range of
SNR. In the low SNR regime, the BHT-based support detector provides
noisy-robust detection against measurement noise. In the high SNR
regime, the DD-structure eliminates quantization error due to the
sampled-message BP from the signal value estimation. In addition, we
applied the knowledge of $x_{min}$ to the proposed algorithm based
on the result of Wainright \emph{et al.}
\cite{Wainwright1},\cite{Wainwright2}. Then, we showed that the use
of $x_{min}$  enables BHT-BP to remove the error-floor of the
MAP-based algorithms, inducing the performance to approach that of
the oracle estimator as SNR increases.

We supported such advantages of BHT-BP via experimental validation.
Our experiments showed that BHT-BP outperforms the other recent NSR
algorithms over entire range of SNR, approaching the recovery
performance of the oracle estimator as SNR increases.

\section*{Appendix I\\Fundamentals of Density-message Passing}
The goal of BP is to iteratively approximate the marginal posterior
densities via a message update rule. The message update rule can
have various forms according to  applications and frameworks. In the
NSR problems, the BP-messages are represented as PDFs since the
signal is a real-valued. We refer to such messages passing as
density-messages passing. In this appendix, we provide the
fundamentals of  the density-message passing under the BP-SM
framework.

In the discussion here, we consider a random linear model
corresponding to \eqref{eq:eq3-4}, given as
\begin{align}\label{eq:eq4-1}
\mathbf{Z}=\mathbf{\Phi}\mathbf{X}+\mathbf{N},
\end{align}
where $\mathbf{X}$ and $\mathbf{Z}$ are random vectors for
$\mathbf{x}_0$ and $\mathbf{z}$ respectively, and $\mathbf{N} \sim
\mathcal{N}(0,\sigma_N^2\mathbf{I})$ is a random vector for the
Gaussian noise vector $\mathbf{n}$. In addition, we assume that the
sensing matrix $\mathbf{\Phi} \in \{0,1\}^{M \times N}$ sufficiently
sparse such that the corresponding bipartite graph is tree-like.

Given $\mathbf{Z}=\mathbf{z}$, we can represent the marginal
posterior density of  $X_i$ in the form of $ {\rm{Posterior =
Prior}} \times \frac{{{\rm{Likelihood}}}}{{{\rm{Evidence}}}}$ using
the Bayesian rule, given as
\begin{align}\label{eq:eq4-2}
f_{X_{i}}(x|\mathbf{Z}=\mathbf{z}) &=   f_{X}(x)\times \frac{ {
f_{\mathbf{Z}}( {{\bf{z}}|X_i=x_i)}}}{f_{\mathbf{Z}}(\mathbf{z})}\\
&\propto f_{X}(x) \times \prod\limits_{j \in N_{\mathcal{V}}(i)}
f_{Z_j}({z}|X_i={x_{i}}),\label{eq:eq4-3}
\end{align}
where  we  postulate that the measurements associated with $X_{i}$,
\emph{i.e.}, $\{ {Z_k}:k \in {N_\mathcal{V}}(i)\} $,  are
statistically independent given $X_{i}=x_i$
\cite{Guo05}-\cite{Guo07} using the tree-like property of
$\mathbf{\Phi}$, to move to \eqref{eq:eq4-3} from \eqref{eq:eq4-2}.

We note each decomposition of the likelihood density, \emph{i.e.},
$f_{Z_j}(z|X_{i}={x_{i}})\,\,\forall j \in N_{\mathcal{V}}(i)$
called \emph{measurement density}, which  is associated with the
marginal posteriors of elements in $\{ {X_k}:k \in
{N_\mathcal{C}}(j),k \neq i\}$. From \eqref{eq:eq4-1}, a measurement
$Z_j$ is represented by
\begin{eqnarray}
\begin{array}{l}\label{eq:th1-2}
{Z_j} = {X_{i}} + \sum\limits_{k \in {N_\mathcal{C}}(j)\backslash \{
i\} } {{X_{k}}}  + {N_j} = {X_{i}} + {Y_j},
\end{array}
\end{eqnarray}
where we define $Y_j:= \sum\limits_{k \in
{N_\mathcal{C}}(j)\backslash \{ i\} } {{X_{k}}}  + {N_j}$. Then, by
factorizing over $Y_j$, the expression of $f_{Z_j}(z|X_{i}=x_i)$
becomes
\begin{align}\label{eq:th1-3}
{f_{{Z_j}}}(z|X_i={x_{i}}) &= {f_{{Z_j}}}(x + y|X_i={x_{i}})\nonumber\\
&= \int\limits_{{Y_j}} {{f_{{Z_j}}}(x + y|{X_{i}}=x_i,{Y_j}) \underbrace {{f_{{Y_j}}}(y|{X_i} = {x_i})}_{={f_{{Y_j}}}(y)}  dy  } \nonumber\\
&= {f_{{Z_j}}}(z|{X_{i}}=x_i,{Y_j}) \otimes {f_{{Y_j}}}( - y),
\end{align}
where ${f_{{Y_j}}}(y|X_{i}=x_i)={f_{{Y_j}}}(y)$ since $Y_j$ is
independent of $X_{i}$. Since  elements in $\{ {X_{k}}: k \in
N_{\mathcal{C}}(j), k \ne i\}$ and $N_j$ are statistically
independent given $\mathbf{Z}=\mathbf{z}$ under the tree-like
assumption \cite{Guo05}-\cite{Guo07}, we may approximately evaluate
the PDF of $Y_j$  by linear convolution as
\begin{eqnarray} \label{eq:th1-5}
{f_{{Y_j}}}(y) =  \left( \bigotimes\limits_{k \in
N_{\mathcal{C}}(j)\backslash\{i\}}
{f_{X_{k}}}(x|\mathbf{Z}=\mathbf{z}) \right) \otimes f_{N_j}(n).
\end{eqnarray}
Finally, by substituting \eqref{eq:th1-5} into \eqref{eq:th1-3} we
obtain the expression of the measurement density given as
\begin{align} \label{eq:th1-1}
&f_{Z_j}(z|X_{i}={x_{i}})=\nonumber\\
&\underbrace{f_{Z_j}(z|X_{i},Y_j)}_{=\delta_{z_j}} \otimes
f_{N_j}(n) \otimes \left(\bigotimes\limits_{k \in
N_{\mathcal{C}}(j)\backslash\{i\}}
{f_{X_{k}}}(-x|\mathbf{Z}=\mathbf{z}) \right),
\end{align}
where $f_{N_j}(n)=f_{N_j}(-n)$ owing to the symmetry shape of
Gaussian densities, and $f_{Z_j}(z|X_{i},Y_j)=\delta_{z_j}$ is true
since no uncertainty on $Z_j$ exists given $X_{i},Y_j$ which contain
information on the neighbored signal elements.

Using the derivation in \eqref{eq:eq4-3} and \eqref{eq:th1-1}, we
can compose the  update rule for the density-message passing. The
practical details  are provided in Section IV-A, on the basis of the
sampled-message approach.

\end{document}